\documentclass[letter,oldversion]{aa} 
\usepackage{graphicx}
\usepackage{aalongtable}
\usepackage{txfonts}
\usepackage{rotating}
\usepackage{lscape}
\usepackage{placeins}
\newcommand{\gsim}{\;\lower.6ex\hbox{$\sim$}\kern-7.75pt\raise.65ex\hbox{$>$}\;}
\newcommand{\lsim}{\;\lower.6ex\hbox{$\sim$}\kern-7.75pt\raise.65ex\hbox{$<$}\;}

\begin{document}
\title{Nature and nurture: the dynamical ages of in situ and accreted globular clusters
in the Milky Way
 }

\author{
Eugenio Carretta\inst{1}
}

\authorrunning{E. Carretta}
\titlerunning{Dynamical ages of globular clusters}

\offprints{E. Carretta, eugenio.carretta@inaf.it}

\institute{
INAF-Osservatorio di Astrofisica e Scienza dello Spazio di Bologna, Via P. Gobetti
 93/3, I-40129 Bologna, Italy}

\date{}

\abstract{The bifurcated age-metallicity relation of globular clusters (GCs) in the Milky
Way (MW) shows that GCs are either originated in situ or accreted into the Galaxy
from former satellites of the MW. The effects of the Galactic tidal field can
leave signatures on the dynamical evolution and structural properties of GCs.
We present a homogeneous census of dynamical ages for a sample of 93 GCs in
the MW, coupled with the knowledge of their common progenitors from the
chemo-dynamical parameters from the Gaia mission, unavailable some years ago.
We found that the majority of accreted GCs (61\%) is dynamically young. This
percentage drops to 38\% for in situ GCs. Excluding the enigmatic low-energy
(LE) GCs, with ambiguous origin, the fraction of dynamically young ex situ GCs
raises to 70\%. A two tail Kolmogorov-Smirnov test shows that the distribution
of dynamical ages of LE GCs cannot be distinguished from the
distribution of in situ bulge and disc GCs. Yet, the LE GCs are firmly located
on the satellite branch of the age-metallicity relation. An explanation may be
that the progenitor of LE GCs plunged very early into the MW, so that the
gravitational field of the MW had time enough to act on the associated GCs.
The dynamical ages offer a statistically robust evidence corroborating the
scenario of an early accretion proposed by several recent studies on the origin
of LE GCs.}

\maketitle

\section{Introduction}

The age-metallicity relation (AMR) of globular clusters (GCs) in the Milky Way
(MW) is bifurcated in two parallel branches (e.g. Leaman et al. 2013, VandenBerg et
al. 2013). The metal-rich branch hosts GCs with disk-like kinematics and consists
of clusters formed in situ in the Galaxy. The metal-poor GCs with
halo-type orbits were suggested to be formed in low mass satellite systems later
accreted, contributing to the hierarchical formation history of the Milky Way.
This double formation channel is strongly substantiated by a wealth of
chemo-dynamical studies exploiting the data from the astrometric Gaia mission
(e.g. Massari et al. 2019, Kruijssen et al. 2019, Forbes 2020, Callingham et al.
2022, Malhan et al. 2022). 

On the other hand, characteristic radii of GCs are used to investigate their
internal dynamics and the effects of the global MW gravitational field on the
long term dynamical evolution of these stellar systems (e.g. Meylan and Heggie
1997, Piatti et al. 2019). In the present letter we present, for a large sample
of MW GCs, a scrutiny of dynamical ages, defined as the ratio between the
chronological ages and the half-mass relaxation times. Since the latter differ
from cluster to cluster depending on the local conditions (e.g. the distance
from the central potential well, the average density of the regions inhabited by
the cluster), different stages of dynamical evolution can be shared by GCs with
the same chronological ages.

We add here the knowledge on the progenitors of different groups of GCs gained
thanks to the Gaia mission, unavailable some years ago, although some
attributions are still uncertain. Most notable is the case of the low energy
(LE) GCs (Massari et al. 2019, hereinafter M19), whose origin is still
attributed to an in situ component or to a former accretion event variously
dubbed as Kraken (Kruijssen et al. 2020: K20), Koala (Forbes 2020), or
Heracles (Horta et al. 2021,  see also Massari et al. 2026).
The combination of these different pieces of information seems to show that
the overwhelming majority of accreted GCs is dynamically young, as opposed to in
situ GCs that show a range in dynamical ages. Moreover, the pattern of dynamical
ages for LE GCs is similar to those of in situ disc and bulge GCs, although LE
GCs are firmly located on the satellite branch of the AMR. 

\section{Dataset and main results}

Following Dalessandro et al. (2024), we computed the ratio $N_h=t_{age}/t_{rh}$
as a proxy of the clusters' dynamical age. The chronological ages $t_{age}$ are from
Kruijssen et al. (2019). While these ages are from the combination of three
other studies, this is currently the most homogeneous set of ages available for 
the largest number of GCs. We adopted the half-mass relaxation time $t_{rh}$
from the publicly available catalogue by Baumgardt et al. (2019) and as in
Dalessandro et al. (2024) we defined dynamically young GCs as those with
$N_h<8$. This is the threshold at which the chemically polluted  population in
GCs, which is the largest stellar component in GCs (e.g. Carretta et al. 2010),
shows a clear change  in the average velocity distribution, moving from a
radially anisotropic velocity distribution to an isotropic pattern in
dynamically old GCs. Dynamical ages are listed in Table~\ref{t:tabapp}.

\begin{figure*}[t]
\centering
\includegraphics[scale=0.23]{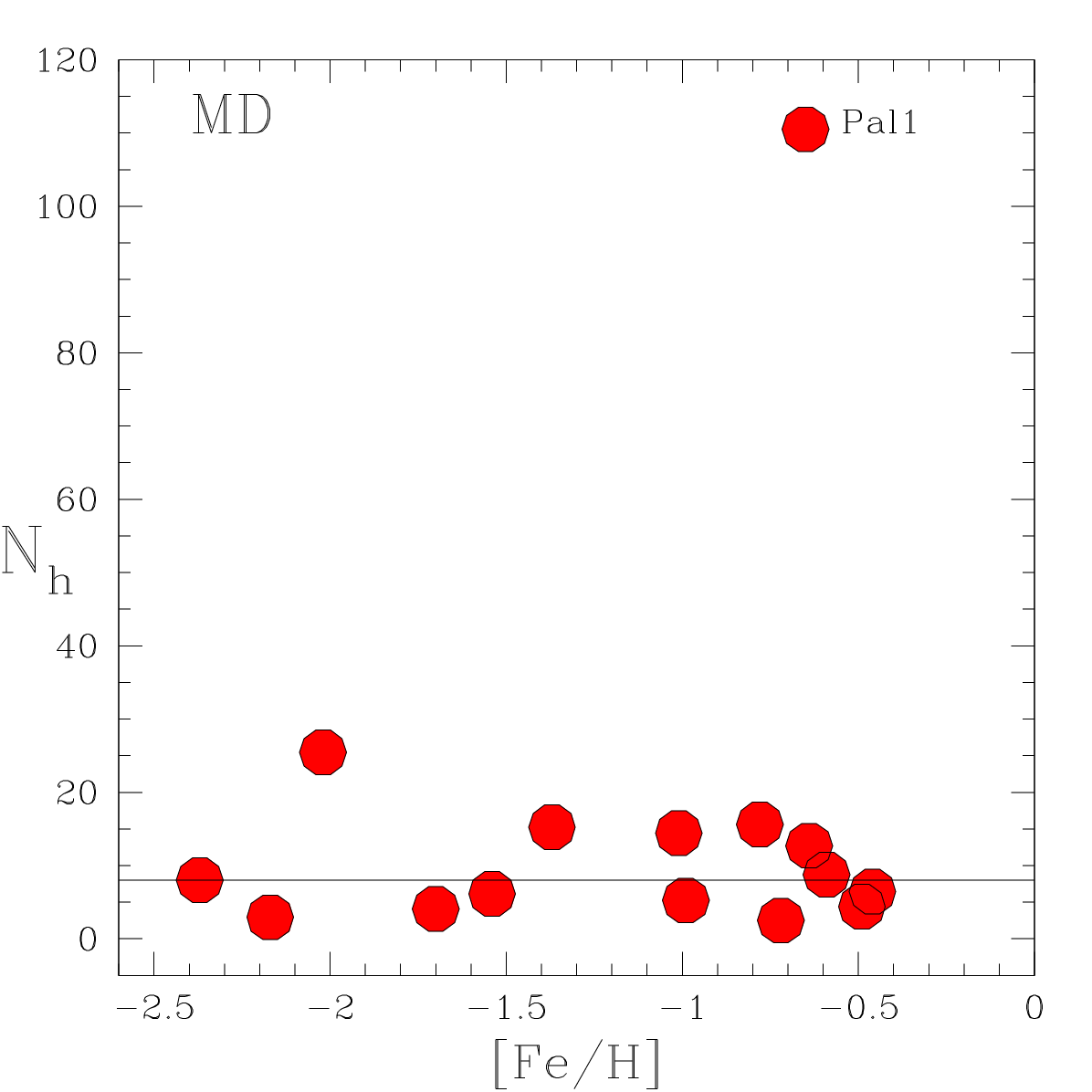}\includegraphics[scale=0.23]{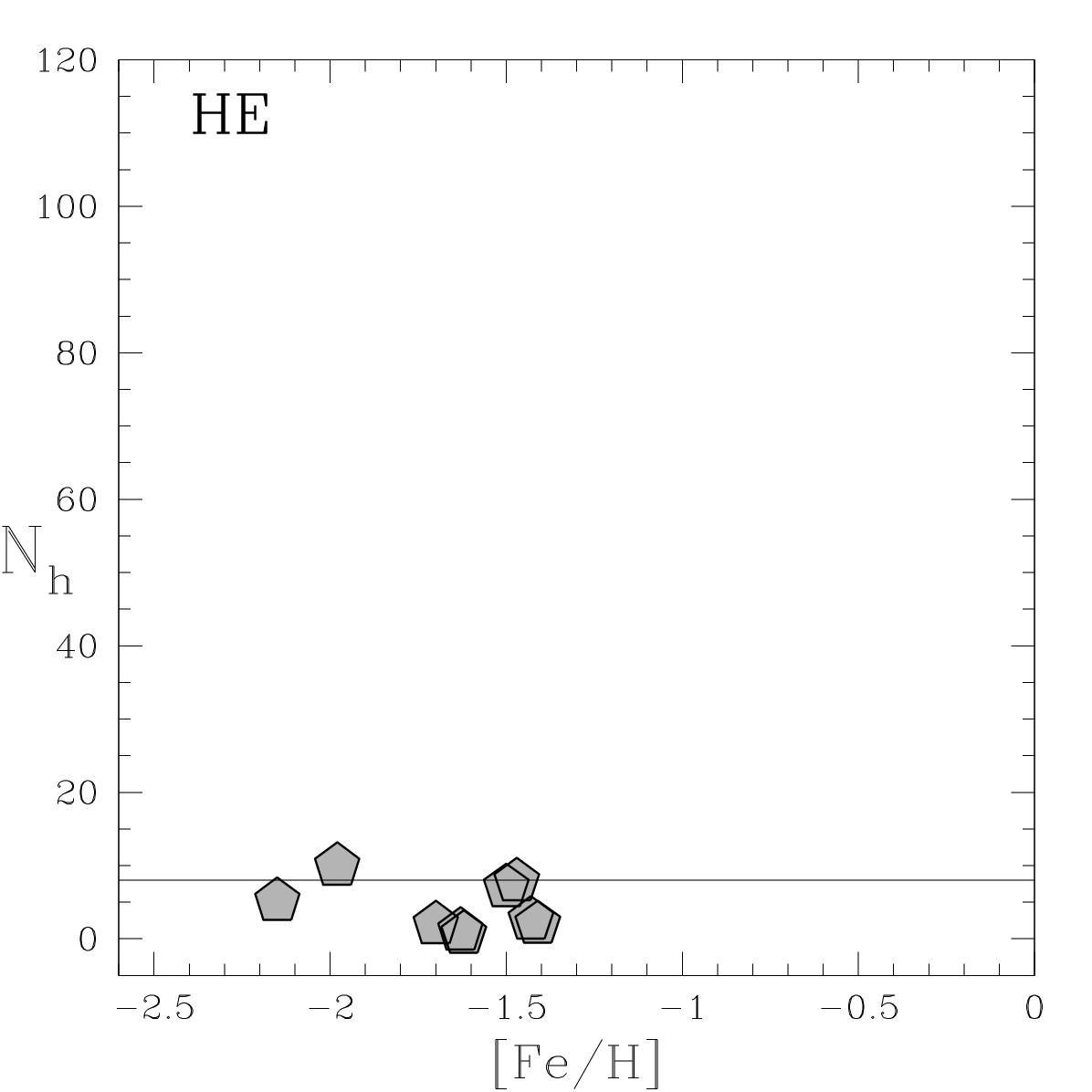}\includegraphics[scale=0.23]{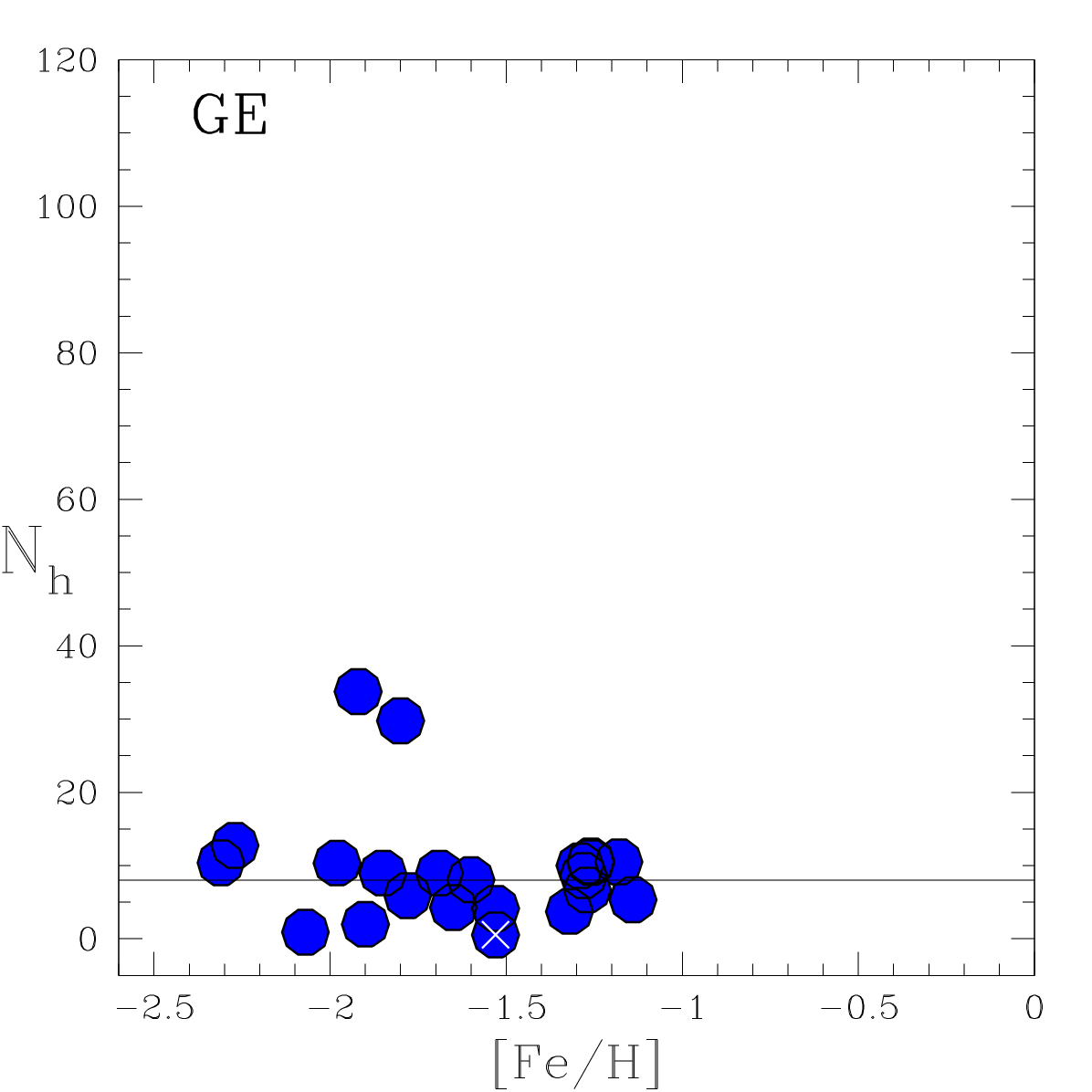}\includegraphics[scale=0.23]{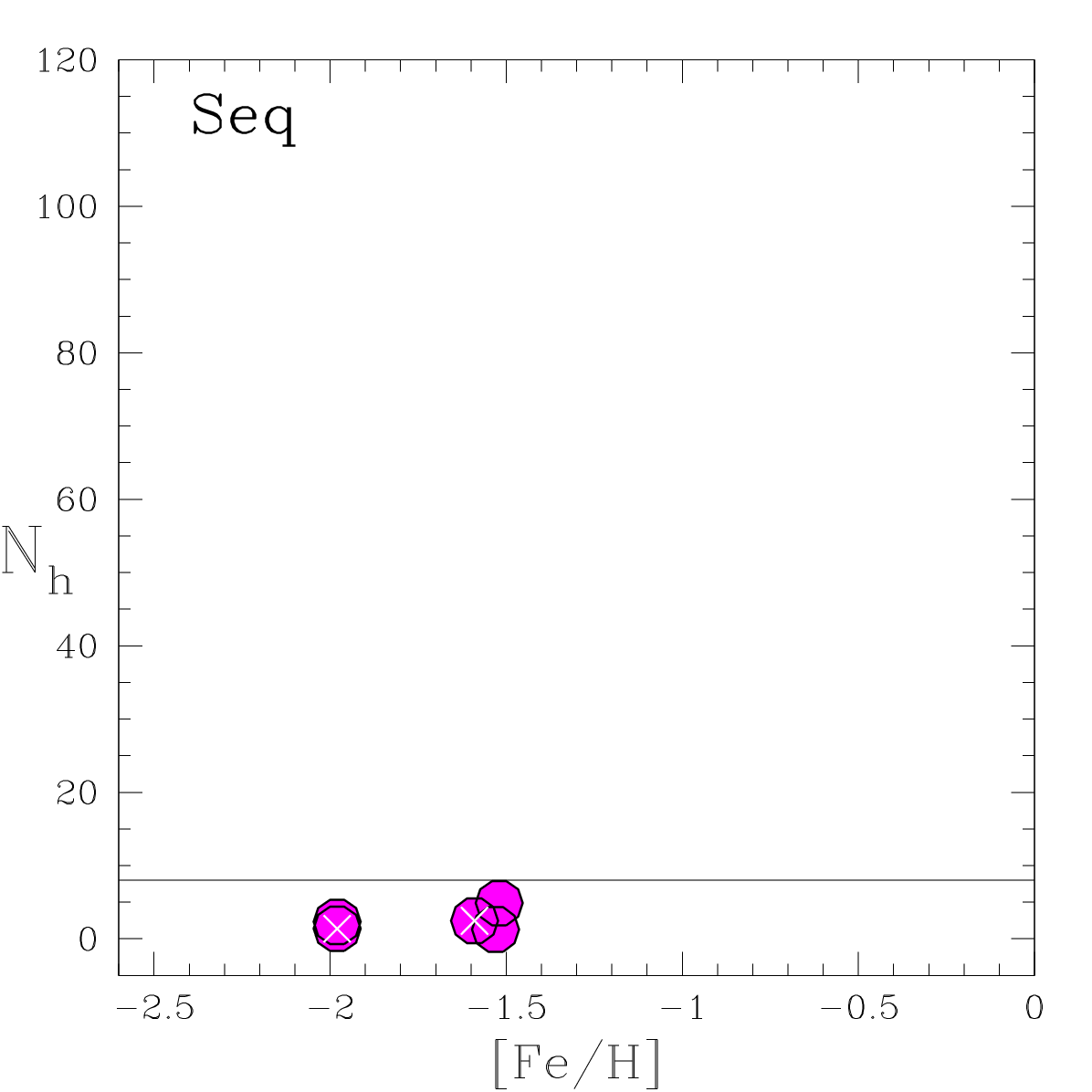}
\includegraphics[scale=0.23]{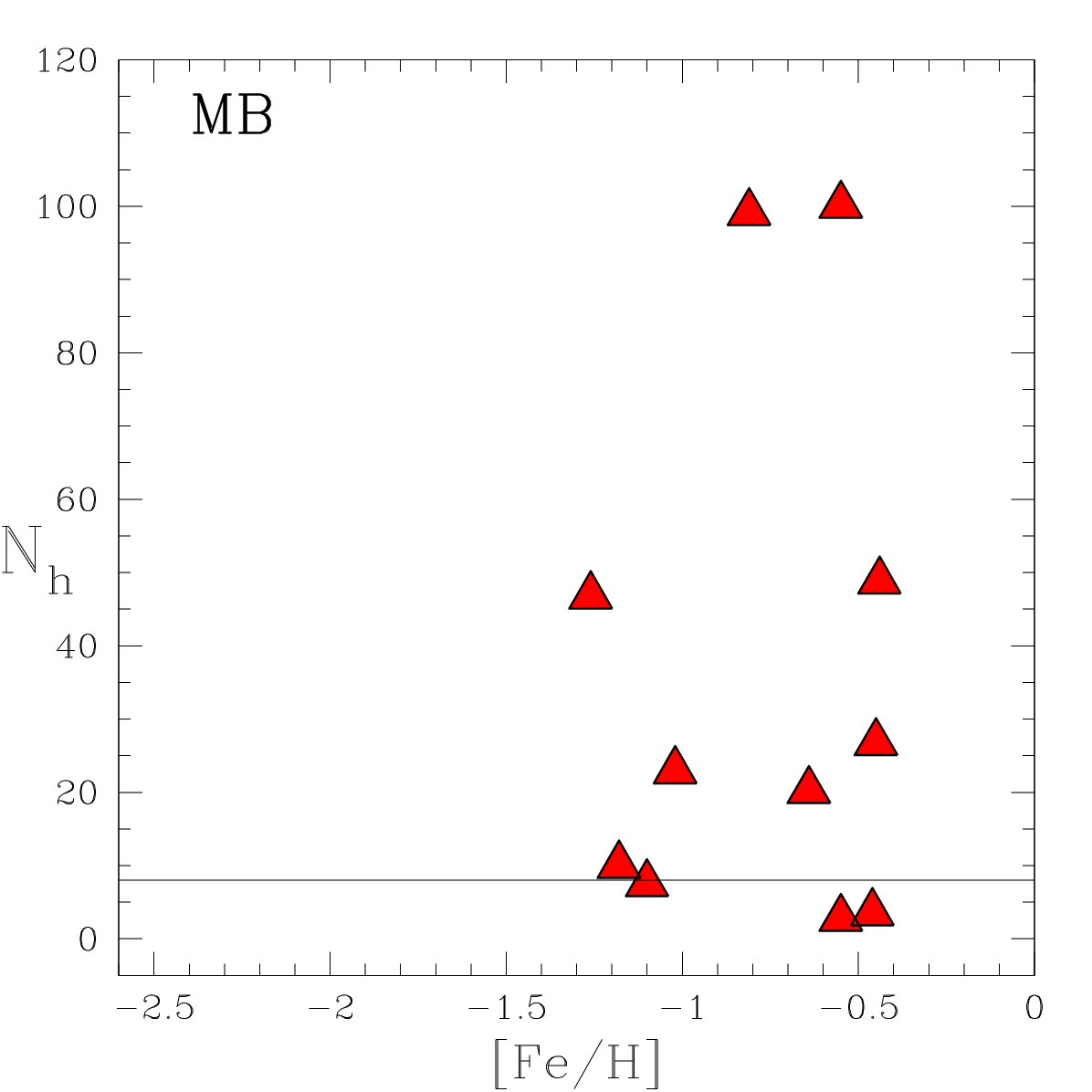}\includegraphics[scale=0.23]{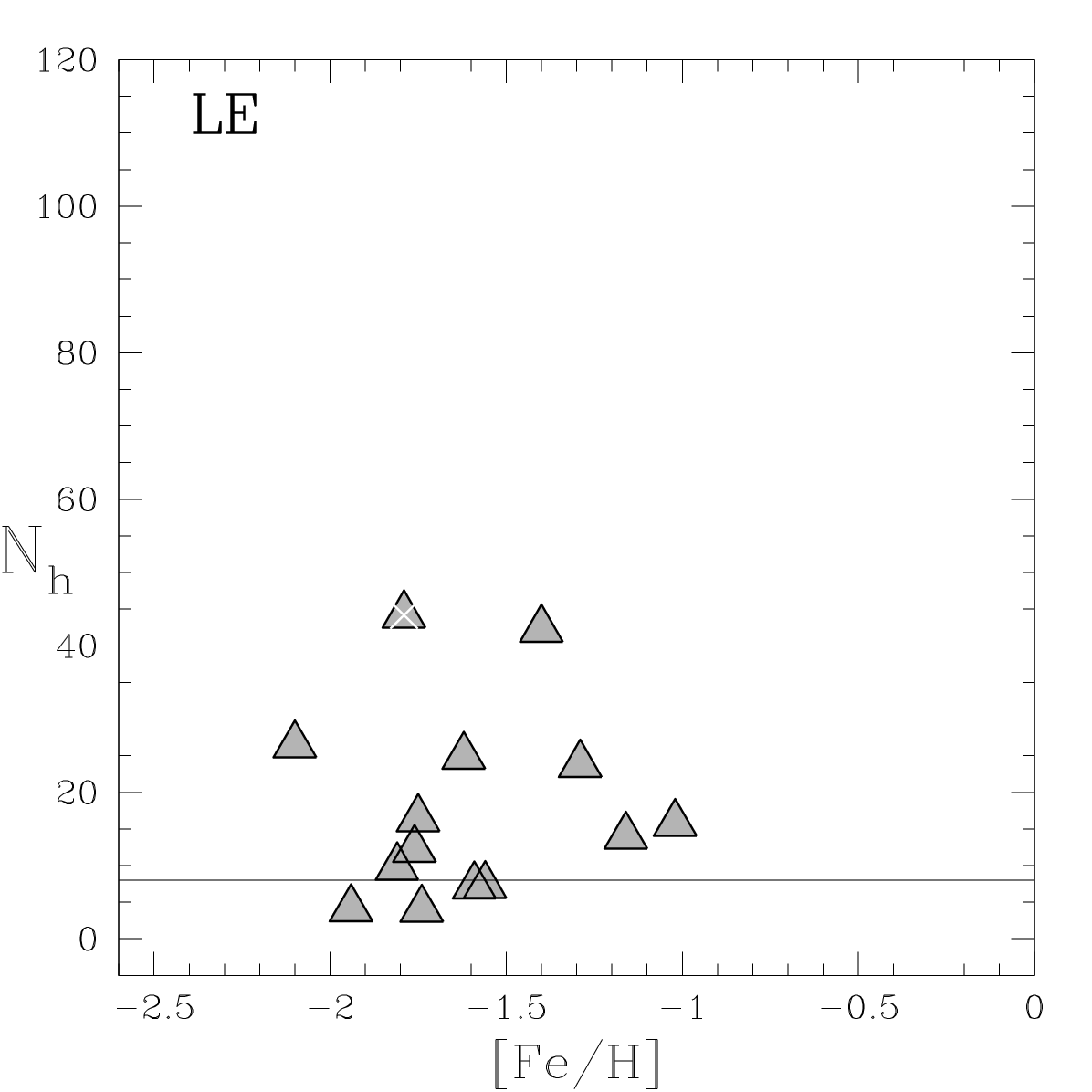}\includegraphics[scale=0.23]{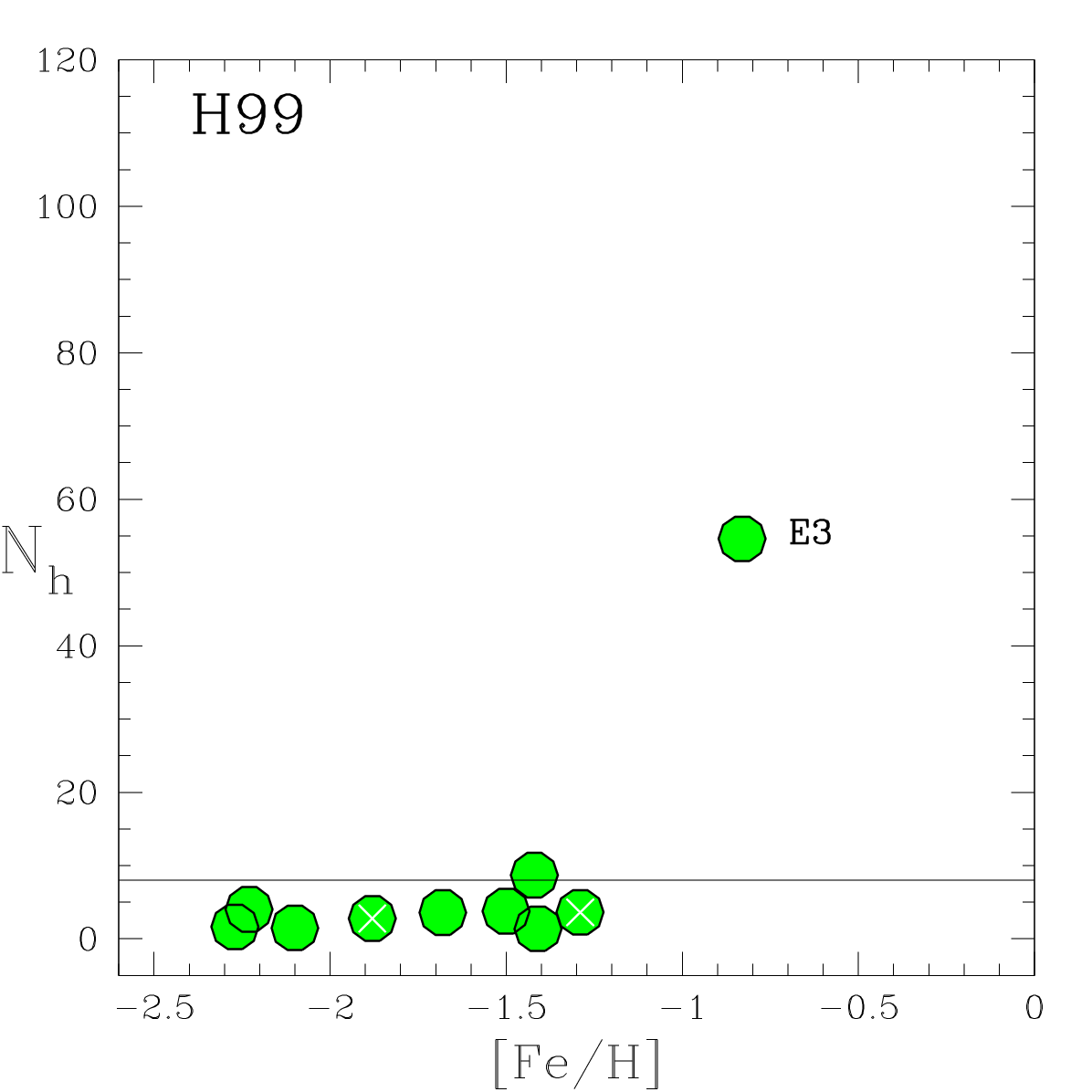}\includegraphics[scale=0.23]{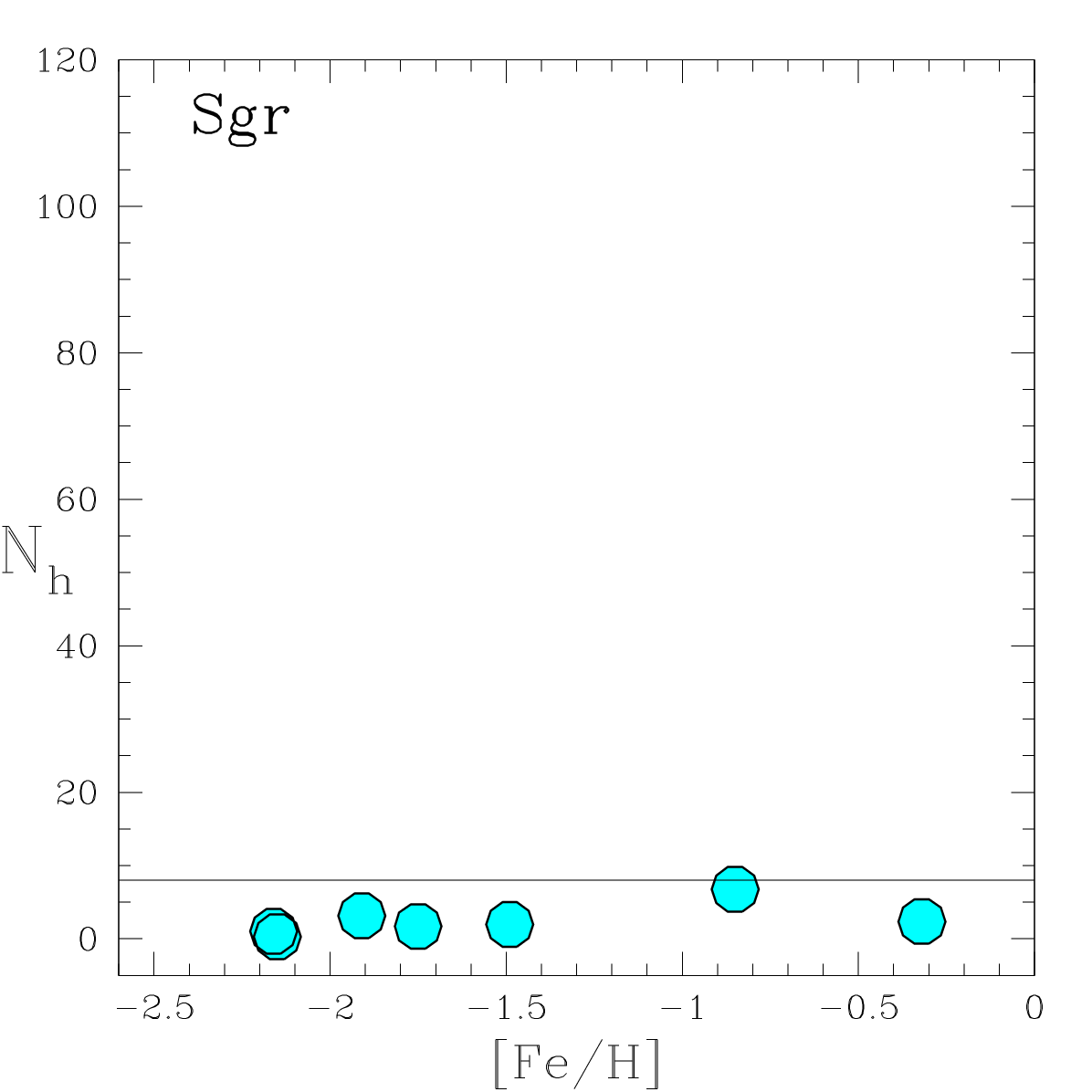}
\caption{The dynamical ages of GCs with different progenitors (different
colours or symbols and labels in each panel) as a function of
the metallicity [Fe/H]. White crosses indicate GCs with uncertain
classification (see text). The line for $N_h=8$ is also shown. Two deviant GCs 
are also indicated in the MD and H99 groups. }
\label{f:nhfe}
\end{figure*}

\begin{figure*}[t]
\centering
\includegraphics[scale=0.18]{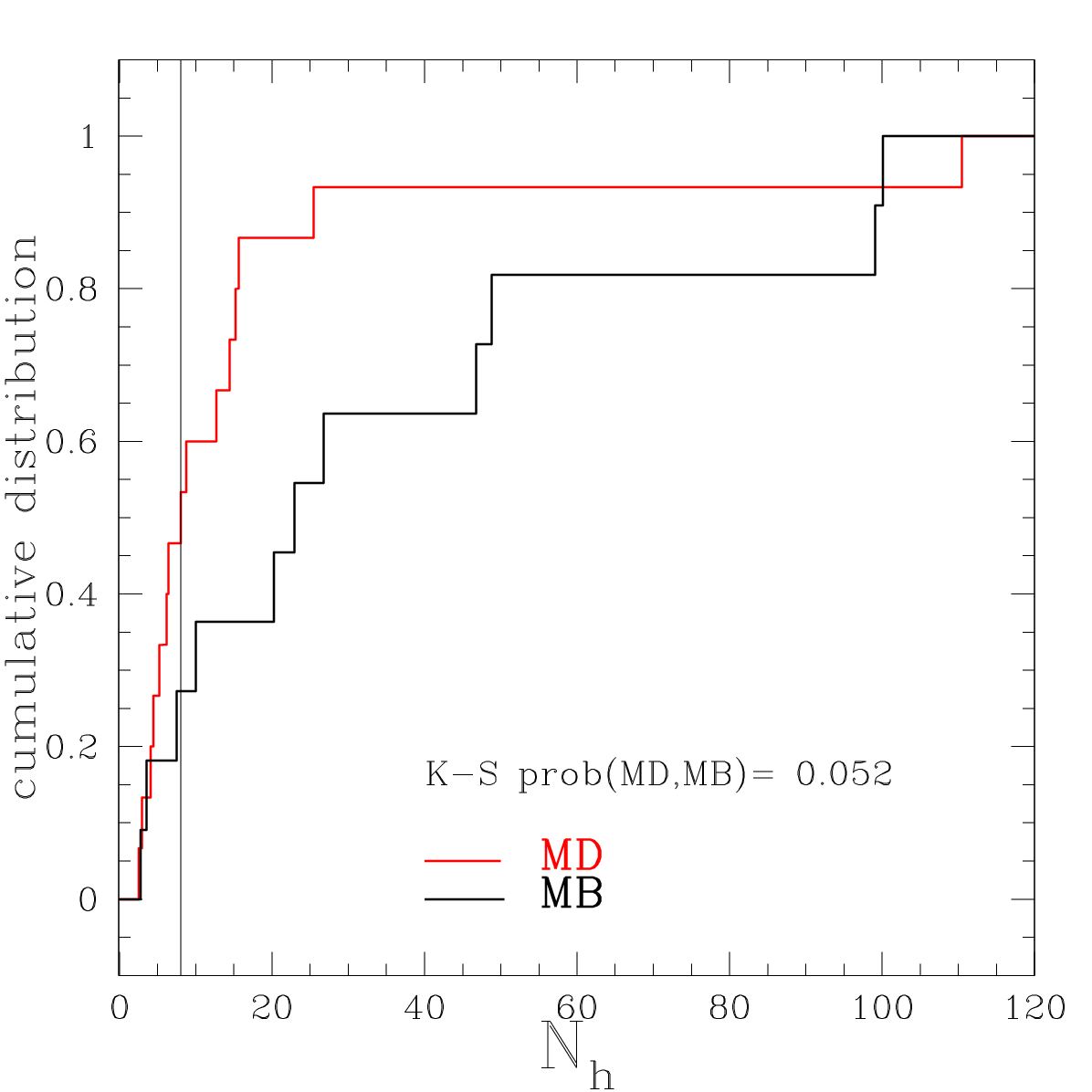}\includegraphics[scale=0.18]{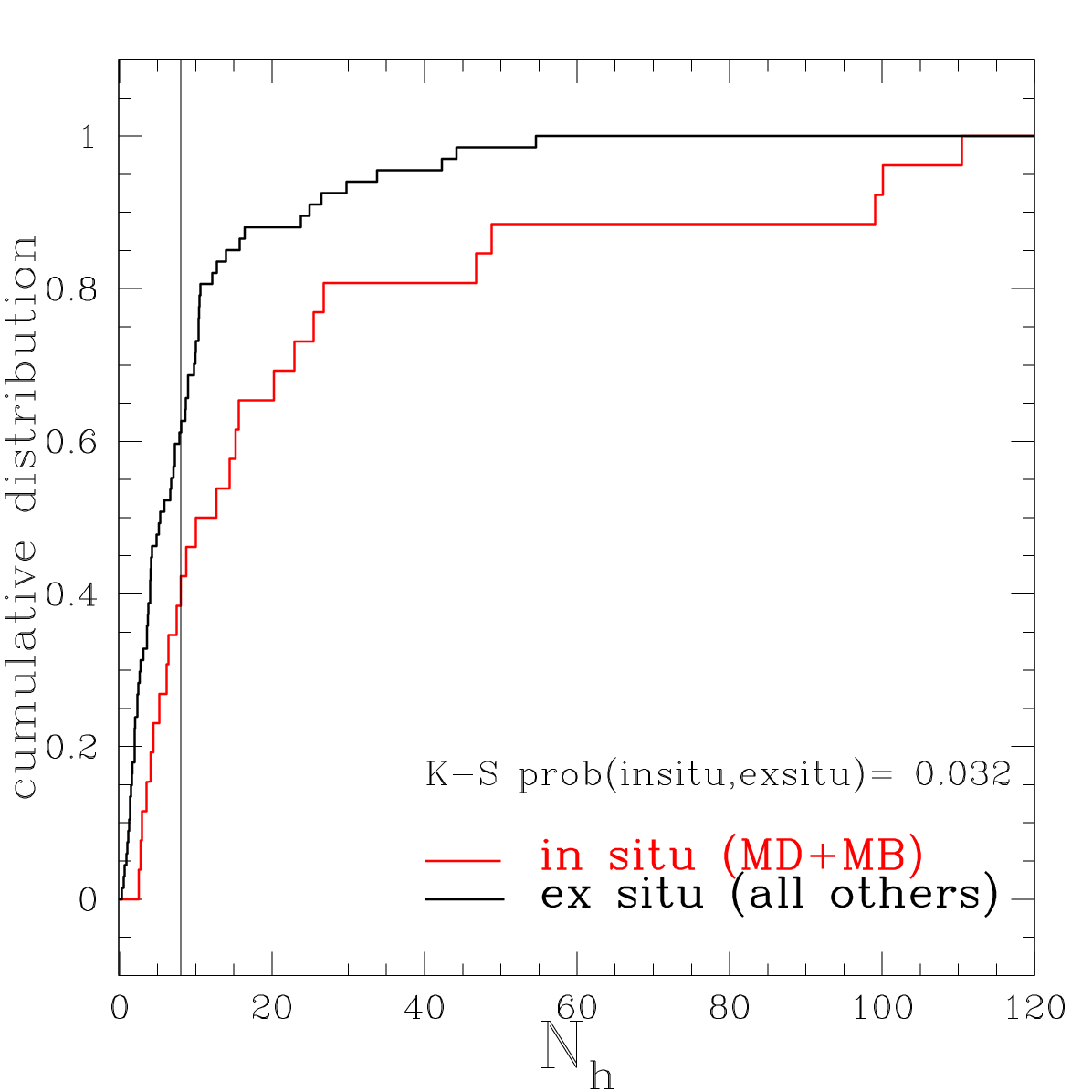}\includegraphics[scale=0.18]{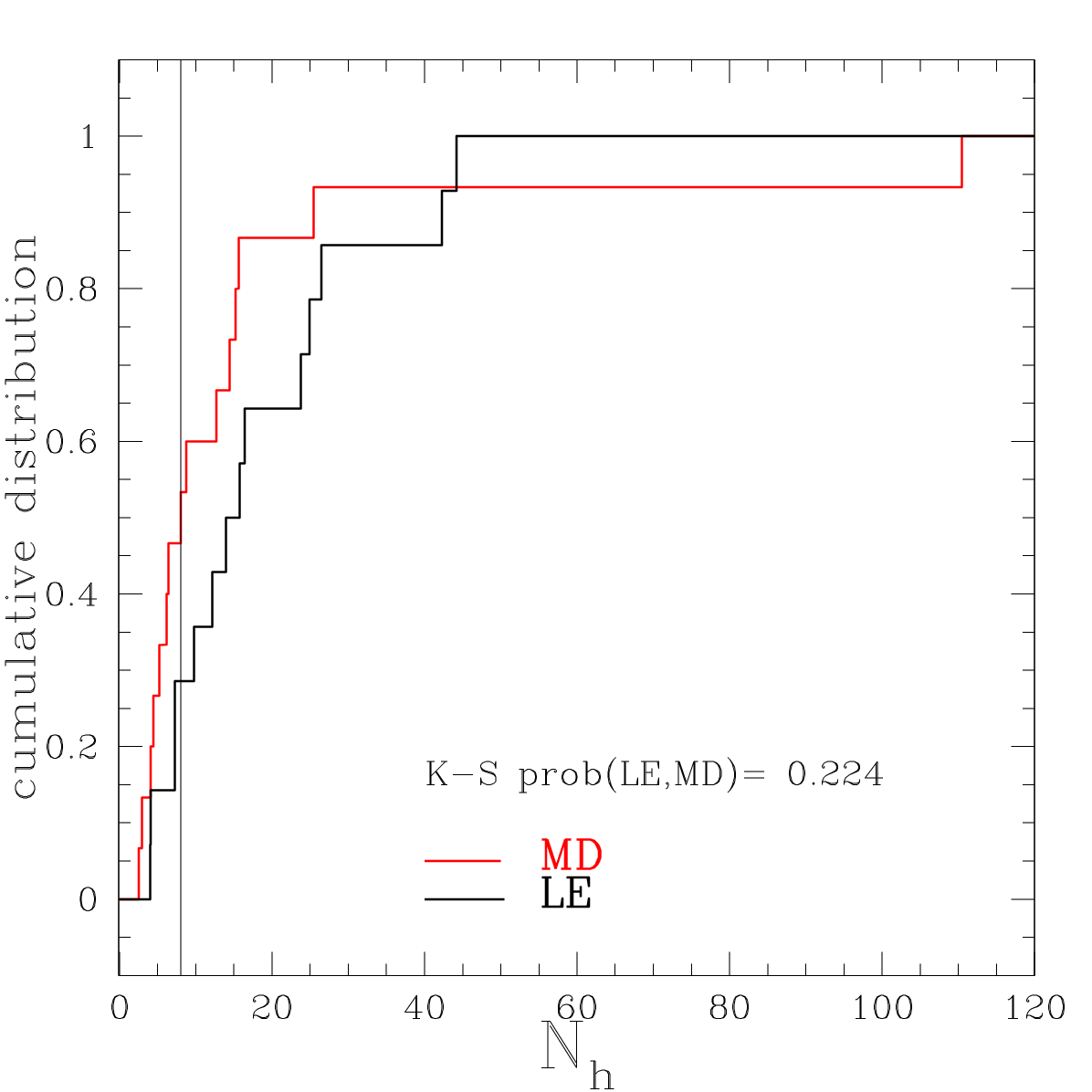}\includegraphics[scale=0.18]{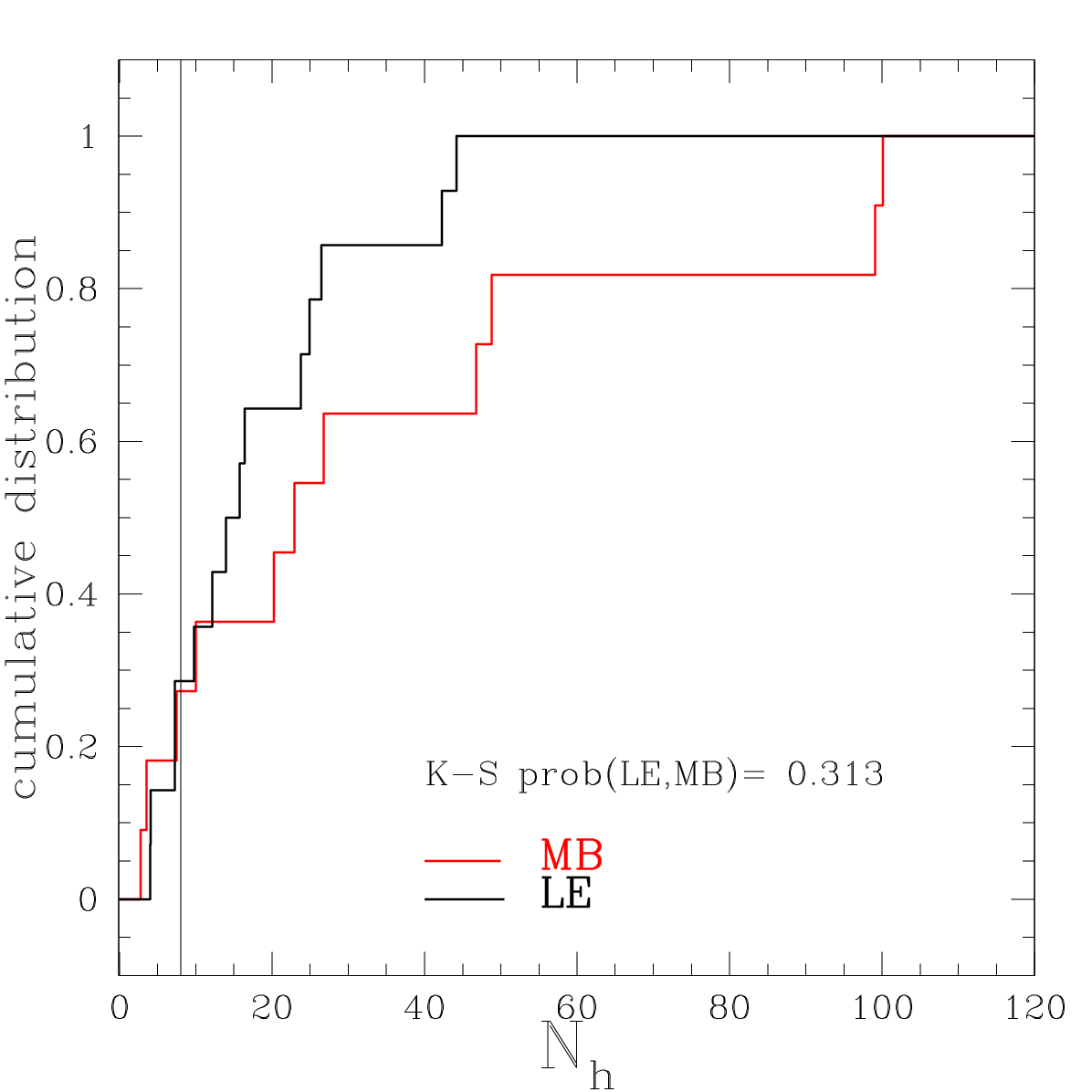}\includegraphics[scale=0.18]{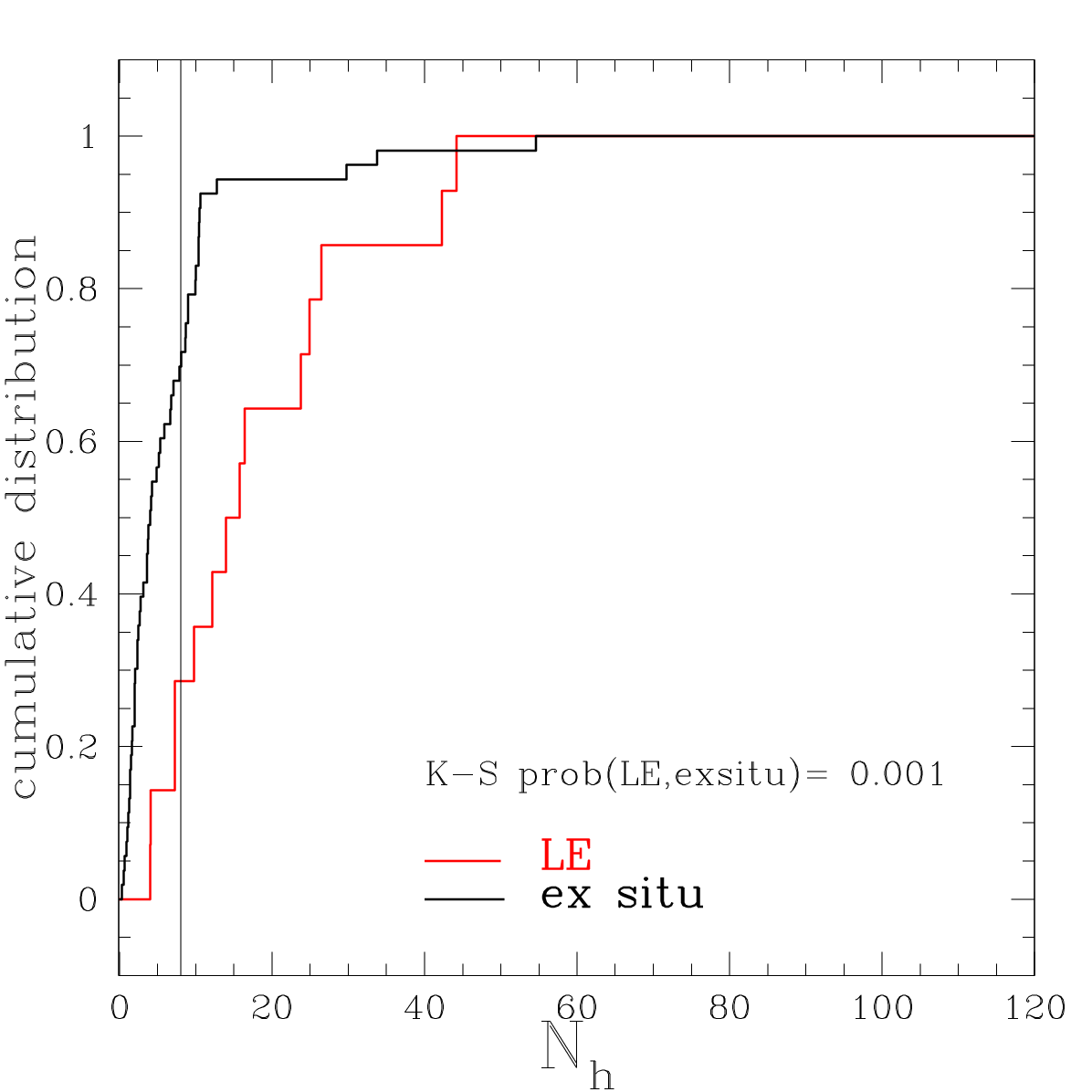}
\caption{Cumulative distributions of dynamical ages for groups of GCs with
different progenitors. In each panel the p-values of the two-sample K-S test are
listed.}
\label{f:ks}
\end{figure*}

The proposed GC membership to different progenitors is adopted mainly from the
classification of M19. For GCs with uncertain progenitors, we adopt the
assignments from others studies, as annotated in Table~\ref{t:tabapp}, where
these GCs are marked with flag=0 and with a white cross on their symbol in the
figures. These ambiguous attributions do not change the main comparison between
in situ and accreted GCs, simply shifting the GCs from a progenitor to another.
In our sample we collected 26 in situ GCs (15 main disc MD, 11 main bulge MB)
and 67 accreted GCs, 22 from Gaia-Enceladus (GE), 10 from the Helmi streams
(H99), 5 from Sequoia (SEQ), 7 from Sagittarius (SGR), 9 from the high energy
group (HE), and 14 from the low energy group (LE).  The only relevant difference
with M19 that may concern the present analysis is  NGC~6441. This is a LE GC in
M19, that we assigned to the MB group, as clearly shown by its chemistry (e.g.
Gratton et al. 2006, Carretta and Bragaglia 2023) and its location on the in
situ branch of the AMR (see Fig.~\ref{f:appfig1}, the discussion in K20, and
Massari et al. 2023).

Our main result is shown in Fig.~\ref{f:nhfe}, where we plot the dynamical ages
of GCs with different progenitors as a function of the metallicity [Fe/H]. The
majority of accreted GCs are dynamically young. All GCs associated to
Sagittarius and Sequoia are below $N_h=8$ like all GCs attributed to the Helmi
streams, except for E3\footnote{Kruijssen et al. (2020) classified this GC as
ambiguous and possibly member of the main Galaxy, excluding it from their
analysis.} and most of the unassigned group of HE GCs with high orbital
energy. A real spread in dynamical ages is shown only among in situ GCs,
with MB GCs presenting the most dynamically evolved GCs. The only GCs on the
satellite branch of the AMR showing a spread of dynamical ages are the
components of the LE group and, to a lesser extent, the GCs associated to
Gaia-Enceladus. We found that $61\pm 10\%$ of the 67 accreted GCs is dynamically
young, where the error is from Poisson statistics. Among the 26 remaining
accreted and dynamically old GCs, a half are members of GE. Exluding LE GCs, the
fraction of dynamically young GCs raises to $70\pm 11\%$. Conversely, the
majority of in situ GCs ($62\pm 15\%$) is composed by dynamically old clusters.

\section{Discussion and conclusions}

We quantified similarities and differences among the GCs with different
progenitors with a two-sample Kolmogorov-Smirnov (K-S) test, under the null
hypothesis that the samples are drawn from the same distribution. The cumulative
distributions of dynamical ages are compared and the resulting p-values are
listed in Table~\ref{t:tableKS}. A few key comparisons are illustrated in
Fig.~\ref{f:ks}. For the groups of GCs in the adopted classification we show in
Fig.~\ref{f:appfig1} and Fig.~\ref{f:appfig2} the AMR and the distribution of
dynamical ages as a function of the Galactocentric distance, age, and total
absolute cluster magnitude (a proxy for the present day mass).

Beginning with in situ GCs and comparing the p-value to the usual $\alpha=0.05$ 
threshold, MB and MD GCs are found not significantly different (see 
Fig.~\ref{f:ks}). However, this similarity seems entirely driven by Pal 1, whose
dynamical age is very old. Excluding Pal 1, the K-S test shows that the
distributions of MB and MD GCs are statistically different (second row in
Table~\ref{t:tableKS}), with a range of dynamical ages in MB GCs twice the range
spanned by MD GCs. We note that K20 claim that Pal 1 cannot be considered as a
genuine disc GC because of its position on the extension of the satellite branch
(see also Fig.~\ref{f:appfig1}) and they omit the cluster from their analysis. 

Unsurprisingly, the distributions of accreted GCs are different from those of in
situ GCs, apart from GCs associated to GE. For these GCs the difference is not
statistically significant. Almost the entirety of GCs associated to the Seq,
Sgr, H99, and HE groups shows dynamically young ages, whereas a range is clearly
apparent among GE GCs.

The most interesting result from Table~\ref{t:tableKS} and Fig.~\ref{f:ks}
concerns the group of LE GCs. The distribution of dynamical ages for these GCs
cannot be distinguished from that of in situ GCs (both disc and bulge sets),
whereas the K-S test indicates that it is extracted from different parent
populations with respect to accreted GCs. 

To better understand dynamical ages we can investigate how the nature (origin)
and the nurture (influence of the tidal field on the dynamical evolution)
affect the ratio $N_h$.
The original nature of GCs is generally indicated well by their location on the
bifurcated AMR. Accreted GCs on the satellite branch generally have lower 
metallicity for a fixed age, and this branch spans a larger range in ages, with
a tail extending to young ages. In situ GCs are mostly located in the branch at
higher metallicity, indicating their formation in a more massive progenitor,
although a tail to lower metallicity is present among MD GCs (see
Fig.~\ref{f:appfig1}).

\begin{figure*}[t]
\centering
\includegraphics[scale=0.30]{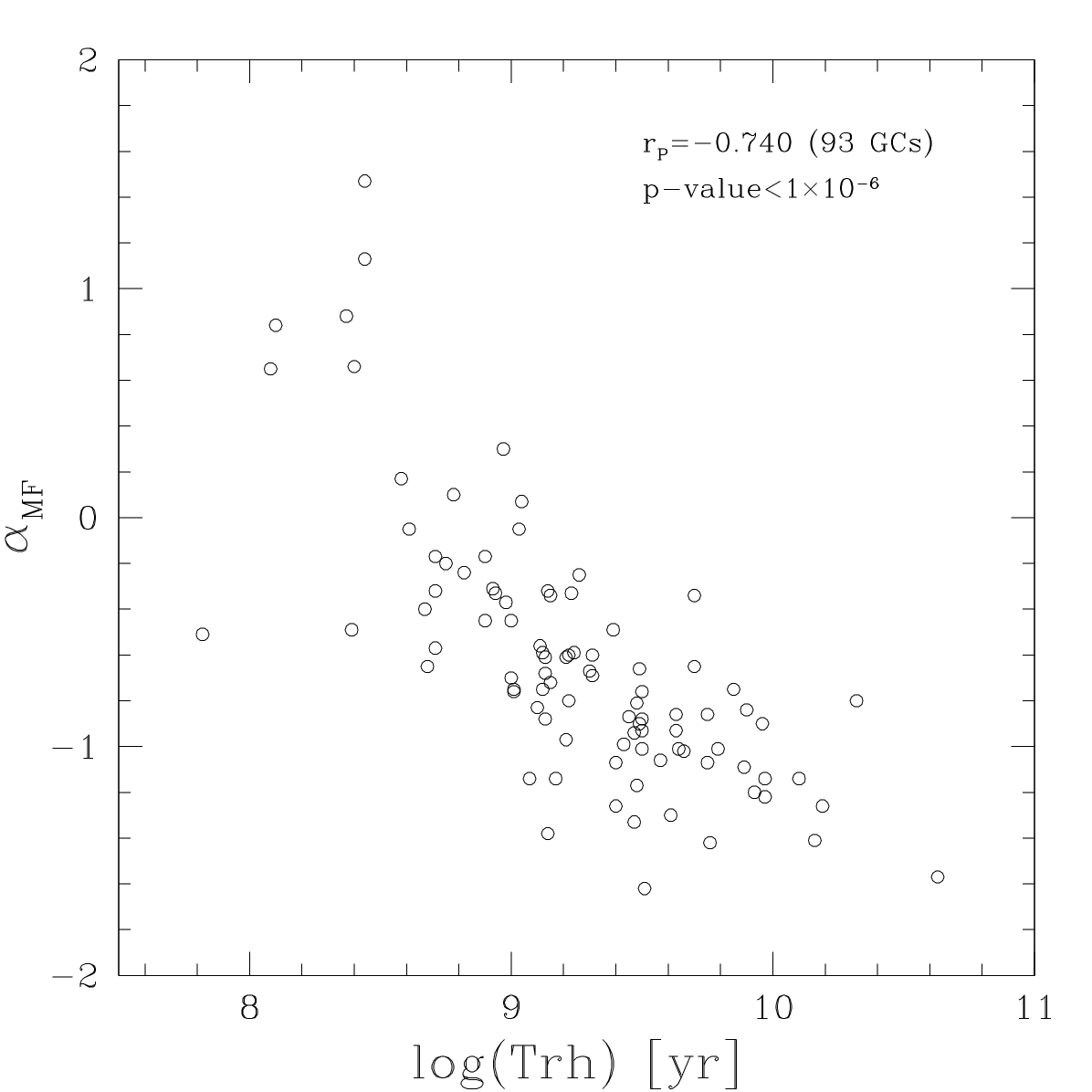}\includegraphics[scale=0.30]{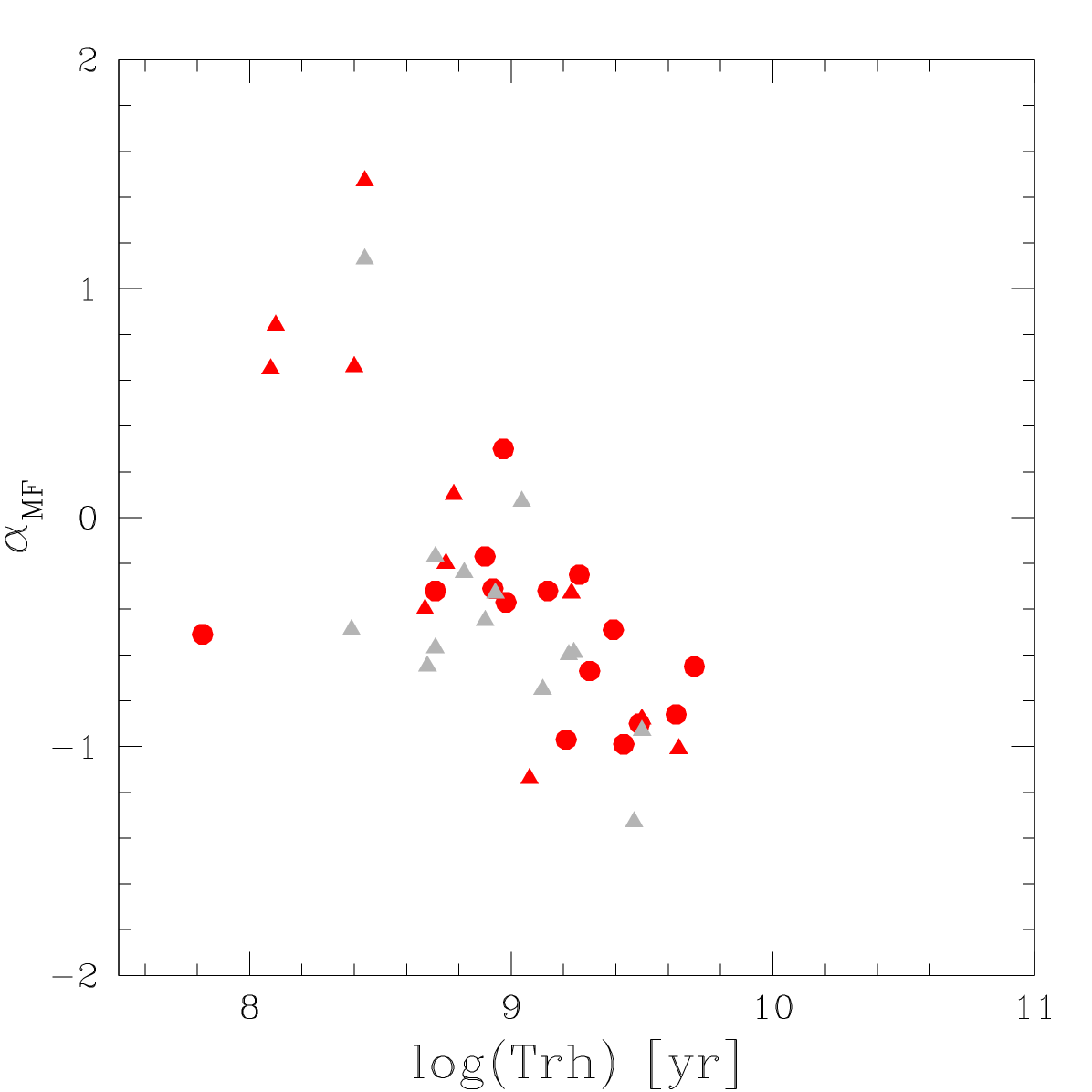}\includegraphics[scale=0.30]{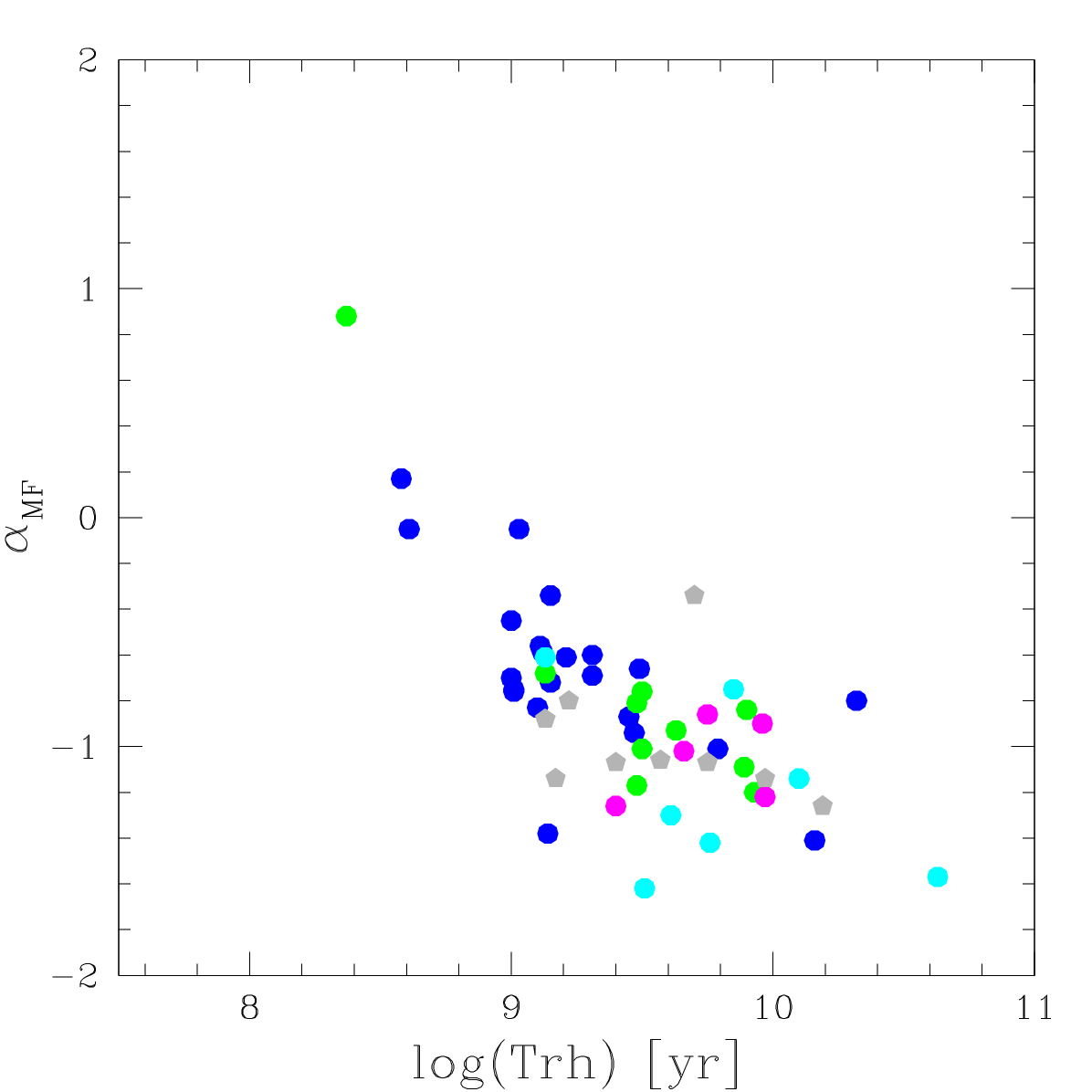}
\caption{Left panel: Slope of the global mass function as a function of the relaxation times
for GC from the catalogue of Baumgardt et al. (2019). In the middle panel only 
MB, MD, and LE GCs, in the right panel only Seq, H99, GE, Sgr, and
HE GCs are plotted. Symbols are as in Fig.~\ref{f:nhfe}.}
\label{f:mf}
\end{figure*}

The effects of the MW tidal field on the dynamical evolution and structural
parameters of GCs are best seen by revisiting the strong correlation between the
mass function slope and the half-mass relaxation times found by Sollima and
Baumgardt (2017). Shown for our sample in the left panel in Fig.~\ref{f:mf},
using data from Baumgardt et al. (2019), this relation is explained by GCs in
stronger tidal field experiencing more mass loss, making them more compact,
reducing their relaxation times and resulting in more positive mass function
slopes (fewer low mass stars) as stars in the outskirts are progressively lost
due to the increasing tidal field. In the middle and right panels of
Fig.~\ref{f:mf} we add the information about the progenitor of GCs unveiling a
sort of segregation onto the two tails of the relation. The effects of the tidal
field are particularly evident among in situ GCs (MD and MB) and LE GCs (middle
panel), whereas the right panel indicates that in accreted GCs the nature of
clusters formed in a weaker gravitational field prevails. The dynamical youth is
apparently maintained for the majority of these GCs also when they become part
of the main Galaxy after the original fragments dissolved. A range of the mass
function slopes suggests that the only group presenting a significant range of
dynamical ages is that of GCs associated to GE (right panel of Fig.~\ref{f:mf},
whereas the restricted range of slopes confirm that in general accreted GCs are
not dynamically highly evolved.

The range of dynamical ages for LE GCs is comparable to that of MB GCs and
larger than that covered by MD GCs. However, the LE GCs are firmly
located on the satellite branch of the AMR (see Fig.~\ref{f:appfig1}).

A possible solution to this conundrum is that the merger event which accreted
the LE GCs in the MW occurred early in the assembly process that formed our
Galaxy as it is now, as suggested by several studies (K20, Horta et al. 2021,
Massari et al. 2026). From TNG50 simulations, Chen and Gnedin (2024) show that
GCs from such a merger have properties virtually indistinguishable from genuine
in situ GCs, since both the contributing and the receiving galactic systems had
comparable mass (and therefore also metallicity) at that epoch. The resulting
merger then allowed the merging GCs to penetrate the inner region in low energy
orbits. In such a manner, the permanence of these GCs in the main Galaxy was
long enough that the effects of the stronger gravitational field were efficient
in shaping the dynamical evolution of these clusters. 

On the other hand, the nature of most accreted GCs is inherent  to their
formation in smaller stellar systems satellites of the main Galaxy, where the
weaker potential well had no way to accelerate the dynamical evolution. 
Apparently, after the accretion these ex situ GCs are deposited further from the
inner regions, typically experience weaker tidal field than the in situ GCs (see
e.g. Meng and Gnedin 2022), and mostly maintain their distinctively young
dynamical ages. 

\begin{figure*}
\centering
\includegraphics[scale=0.30]{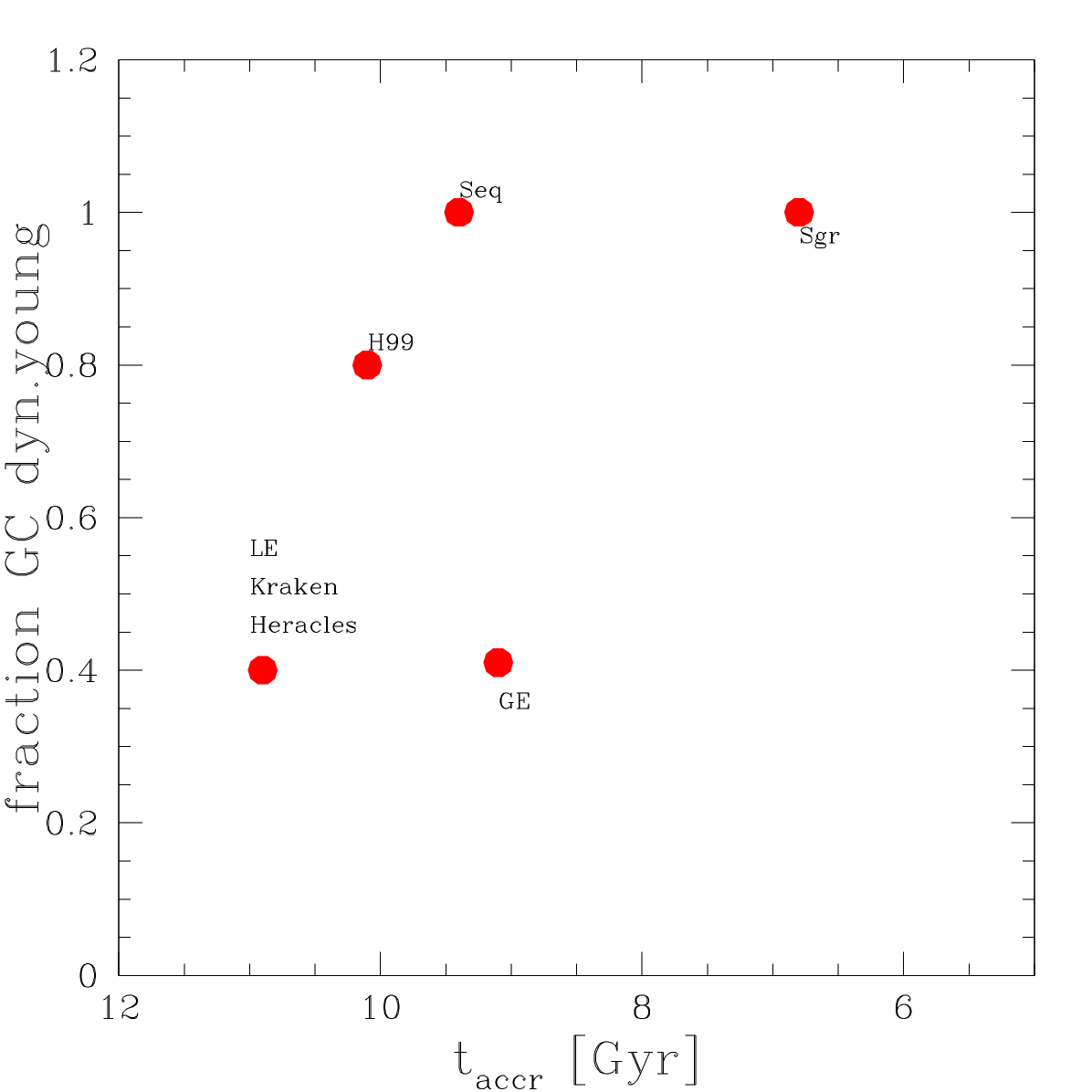}\includegraphics[scale=0.30]{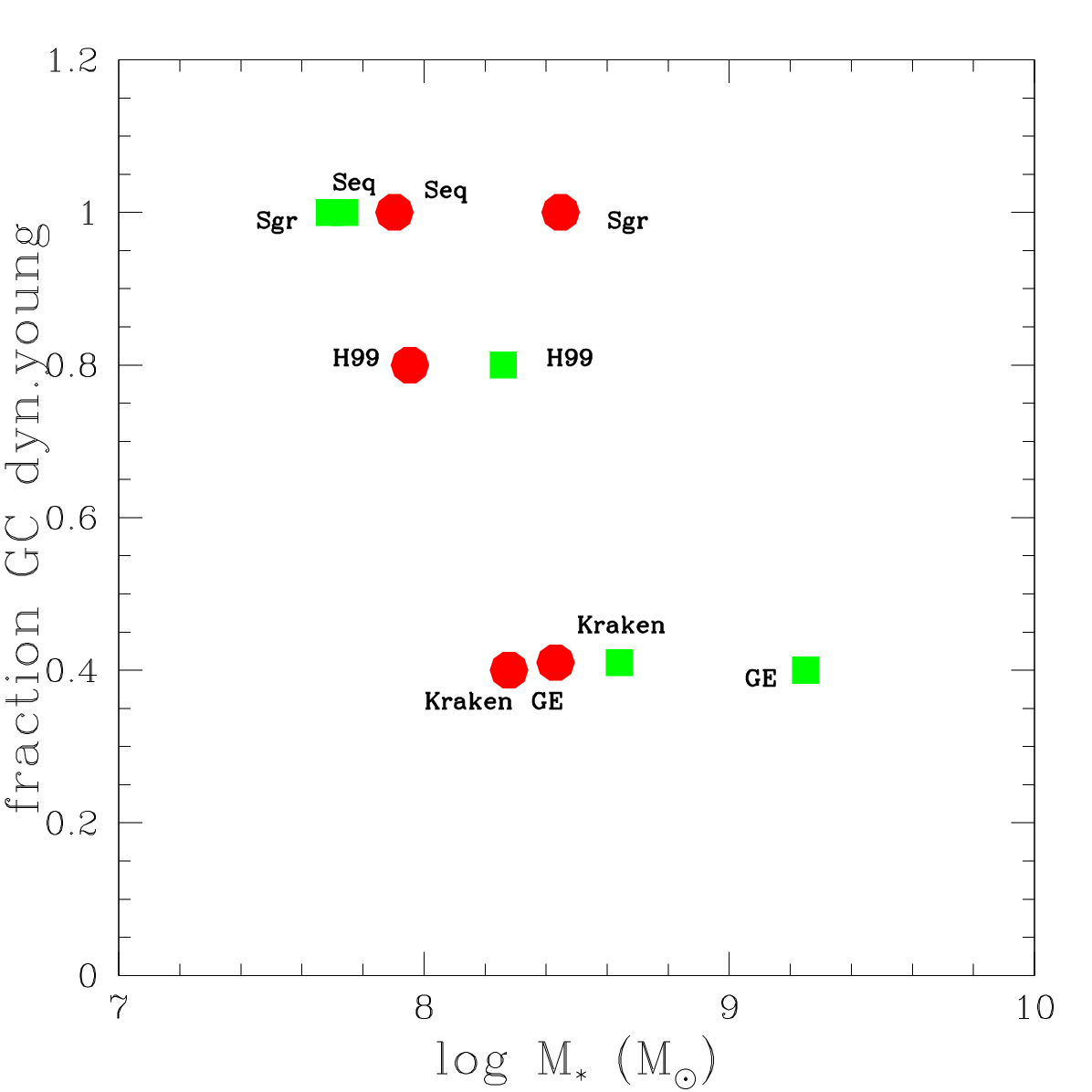}
\caption{Left panel: Fraction of dynamically young GCs associated to different progenitors
as a function of the time of accretion in the MW estimated by K20. Right panel:
the fraction versus the stellar mass of the progenitors from K20 (red circles)
and from the unbiased estimates of Callingham et al. (2022: green squares).}
\label{f:rat}
\end{figure*}

If the interplay of nature and nurture is shaping the dynamical ages of GCs in
the MW, we should find some correspondence with the global properties of the
progenitor systems. Some evidence is shown in Fig.~\ref{f:rat}, where in the
left panel we plot the fraction of GCs we found dynamically young attributed to
each progenitor system as a function of the time of accretion estimated in K20.
While the relation may be not very tight and the uncertainties depend on the
proposed membership of GCs, this figure shows that the fraction of dynamically
young GCs is lower for those associated to merger events that occurred early in
the assembly history of the MW. This fraction increases as the estimated stellar
mass of the progenitors decreases, as expected if the tidal field of the
original galaxy had a lesser impact on the clusters' dynamical evolution. This
is shown in the right panel of Fig.~\ref{f:rat} with mass values from K20 and
from the unbiased estimates from Callingham et al. (2022). In the latter case,
the linear regression is statistically significant (p-value=0.016) even with
only 5 data points.

\section{Summary}

We present a homogeneous census of dynamical ages for a set of 93 GCs in the 
MW. We found that the majority of GCs located on the satellite branch of the AMR
is dynamically young, whereas most in situ GCs seem to be more dynamically
evolved. In summary, the dynamical ages of GCs coupled to the knowledge of their
original progenitors can help a better understanding of their properties. The
nature of accreted GCs, formed in a weaker gravitational field, is revealed by
their dynamically young ages. The scarce evidence of dynamical evolution seems
maintained also after the merger of their progenitors with the main Galactic
body. The only exception is the group of LE GCs, possibly originated in a
fragment that was accreted into the MW at very early epochs. The GCs associated
to Gaia-Enceladus appear as a transition group, with a mixture of dynamically
old and young ages.  While several conclusions of this scrutiny are not
unexpected, homogeneous sets of dynamical ages may be another  observable
property to be used in the clustering of GCs to reconstruct their origin and the
assembly history of the MW.

\begin{table}
\centering
\caption{Two sample K-S results}
\begin{tabular}{lr|lr}
\hline

samples & p(K-S) & samples & p(K-S) \\
\hline

MD - MB                   & 0.052 & in situ - ex situ$^2$     & 0.002 \\
MD$^1$ - MB               & 0.023 & LE - GE		       & 0.015 \\
in situ - GE              & 0.056 & LE - H99		       & 0.000 \\
in situ - H99             & 0.002 & LE - Seq		       & 0.003 \\
in situ - Seq             & 0.004 & LE - Sgr		       & 0.001 \\
in situ - Sgr             & 0.002 & LE - HE		       & 0.011 \\
in situ - LE              & 0.800 & LE - MD		       & 0.224 \\
in situ - HE              & 0.025 & LE - MB		       & 0.313 \\
in situ - ex situ         & 0.032 & LE - ex situ	       & 0.001 \\

\hline
\end{tabular}

\tablefoot{
1-without Pal 1; 2-without LE GCs
}
\label{t:tableKS}

\end{table}

\begin{acknowledgements}
This research has made large use of the SIMBAD database (in particular  Vizier),
operated at CDS, Strasbourg, France, and of the NASA's Astrophysical Data 
System. I warmly thank Angela Bragaglia for several valuable suggestions.
\end{acknowledgements}

\begin{appendix}
\onecolumn

\section{Dynamical ages and progenitors of GCs}

\begin{table*}[h]
\centering
\caption{Classification by progenitor}
\begin{tabular}{llr|llr|llr|}
\hline

GC & orig & flag & GC & orig & flag & GC & orig & flag\\
\hline

  NGC104 &  MD & 1 &	Pal5 &  H99 & 1 & NGC6496 &  MD & 1 \\
  NGC288 &  GE & 1 & NGC5897 &   GE & 1 & NGC6535 &  LE & 0 \\
  NGC362 &  GE & 1 & NGC5904 &  H99 & 0 & NGC6544 &  LE & 1 \\
 NGC1261 &  GE & 1 & NGC5927 &   MD & 1 & NGC6541 &  LE & 1 \\
    Pal1 &  MD & 1 & NGC5946 &   LE & 1 & NGC6584 &  HE & 1 \\
     AM1 &  HE & 1 & NGC5986 &   LE & 1 & NGC6624 &  MB & 1 \\
Eridanus &  HE & 1 &   Pal14 &   HE & 1 & NGC6637 &  MB & 1 \\
 NGC1851 &  GE & 1 & NGC6093 &   LE & 1 & NGC6652 &  MB & 1 \\
 NGC1904 &  GE & 1 & NGC6101 &  SEQ & 0 & NGC6656 &  MD & 1 \\
 NGC2298 &  GE & 1 & NGC6121 &   LE & 1 & NGC6681 &  LE & 1 \\
 NGC2419 & SGR & 1 & NGC6144 &   LE & 1 & NGC6712 &  LE & 1 \\
 NGC2808 &  GE & 1 & NGC6171 &   MB & 1 & NGC6715 & SGR & 1 \\
      E3 & H99 & 1 & NGC6205 &   GE & 1 & NGC6717 &  MB & 1 \\
    Pal3 &  HE & 1 & NGC6218 &   MD & 1 & NGC6723 &  MB & 1 \\
 NGC3201 & SEQ & 0 & NGC6235 &   GE & 1 & NGC6752 &  MD & 1 \\
    Pal4 &  HE & 1 & NGC6254 &   LE & 1 & NGC6779 &  GE & 1 \\
 NGC4147 &  GE & 1 &   Pal15 &   GE & 1 &    Arp2 & SGR & 1 \\
 NGC4372 &  MD & 1 & NGC6266 &   MB & 1 & NGC6809 &  LE & 1 \\
  Rup106 & H99 & 1 & NGC6273 &   LE & 1 &    Ter8 & SGR & 1 \\
 NGC4590 & H99 & 1 & NGC6284 &   GE & 1 & NGC6838 &  MD & 1 \\
 NGC4833 &  GE & 1 & NGC6287 &   LE & 1 & NGC6864 &  GE & 1 \\
 NGC5024 & H99 & 1 & NGC6304 &   MB & 1 & NGC6934 &  HE & 1 \\
 NGC5053 & H99 & 1 & NGC6341 &   GE & 1 & NGC6981 & H99 & 1 \\
 NGC5139 &  GE & 0 & NGC6342 &   MB & 1 & NGC7006 & SEQ & 1 \\
 NGC5272 & H99 & 1 & NGC6352 &   MD & 1 & NGC7078 &  MD & 1 \\
 NGC5286 &  GE & 1 & NGC6366 &   MD & 1 & NGC7089 &  GE & 1 \\
 NGC5466 & SEQ & 1 & NGC6362 &   MD & 1 & NGC7099 &  GE & 1 \\
 NGC5634 & H99 & 0 & NGC6388 &   MB & 1 &   Pal12 & SGR & 1 \\
 NGC5694 &  HE & 1 & NGC6397 &   MD & 1 & NGC7492 &  GE & 1 \\
  IC4499 & SEQ & 1 & NGC6426 &   HE & 1 &    Ter7 & SGR & 1 \\
 NGC5824 & SGR & 1 & NGC6441 &   MB & 1 &  Lynga7 &  MD & 1 \\

\hline
\end{tabular}

\tablefoot{
Flag=0 indicates GCs with uncertain attribution: NGC~3201 (SEQ in Forbes 2020,
Callingham et al. 2022, Kruijssen et al. 2020, Horta et al. 2021, Youakim and
Lind 2025); NGC~5139 (GE in Callingham et al. 2022); NGC~5634 (H99 in Forbes
2020, Callingham et al. 2022, Youakim and Lind 2025); NGC~5904 (H99 in Forbes
2020, Callingham et al. 2022); NGC~6101 (SEQ in Forbes 2020, Callingham et al.
2022, Malhan et al. 2022, Youakim and Lind 2025); NGC~6535 (LE in Massari 2025;
controversial assignment in other studies). Other particular cases: Pal1 is on
the satellite branch in Kruijssen et al. (2020) who omit it from the analysis;
NGC~6441 is possibly a main progenitor GC according to Kruijssen et al. (2020),
E3 is omitted from the analysis in Kruijssen et al. (2020).
}

\label{t:tabapp}

\end{table*}

\section{Dynamical ages as a function of various parameters}

We show the AMR of GCs with different progenitors and their dynamical ages as a
function of global parameters like Galactocentric distance, total absolute 
magnitude and age.

\begin{figure*}[h]
\centering
\includegraphics[scale=0.40]{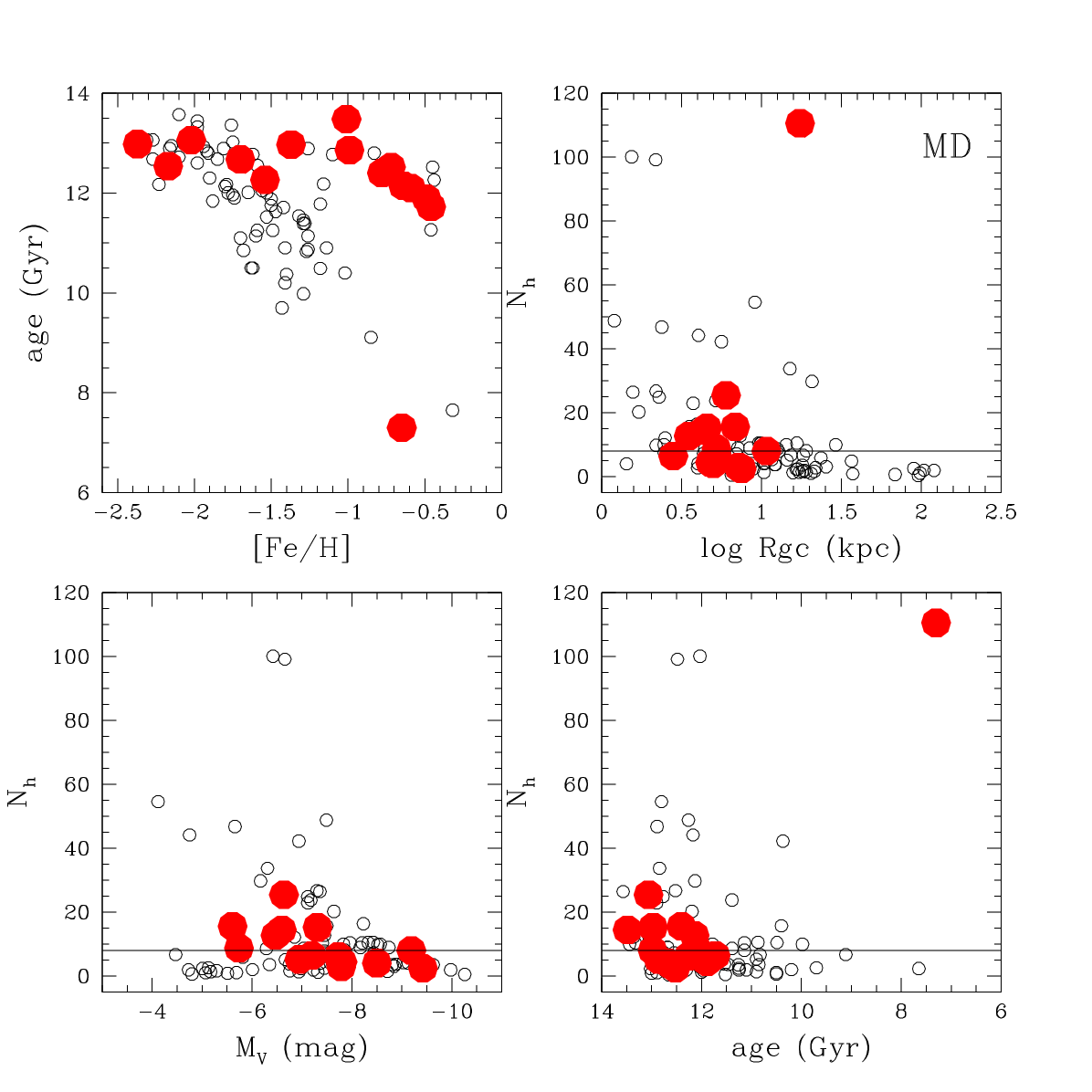}\includegraphics[scale=0.40]{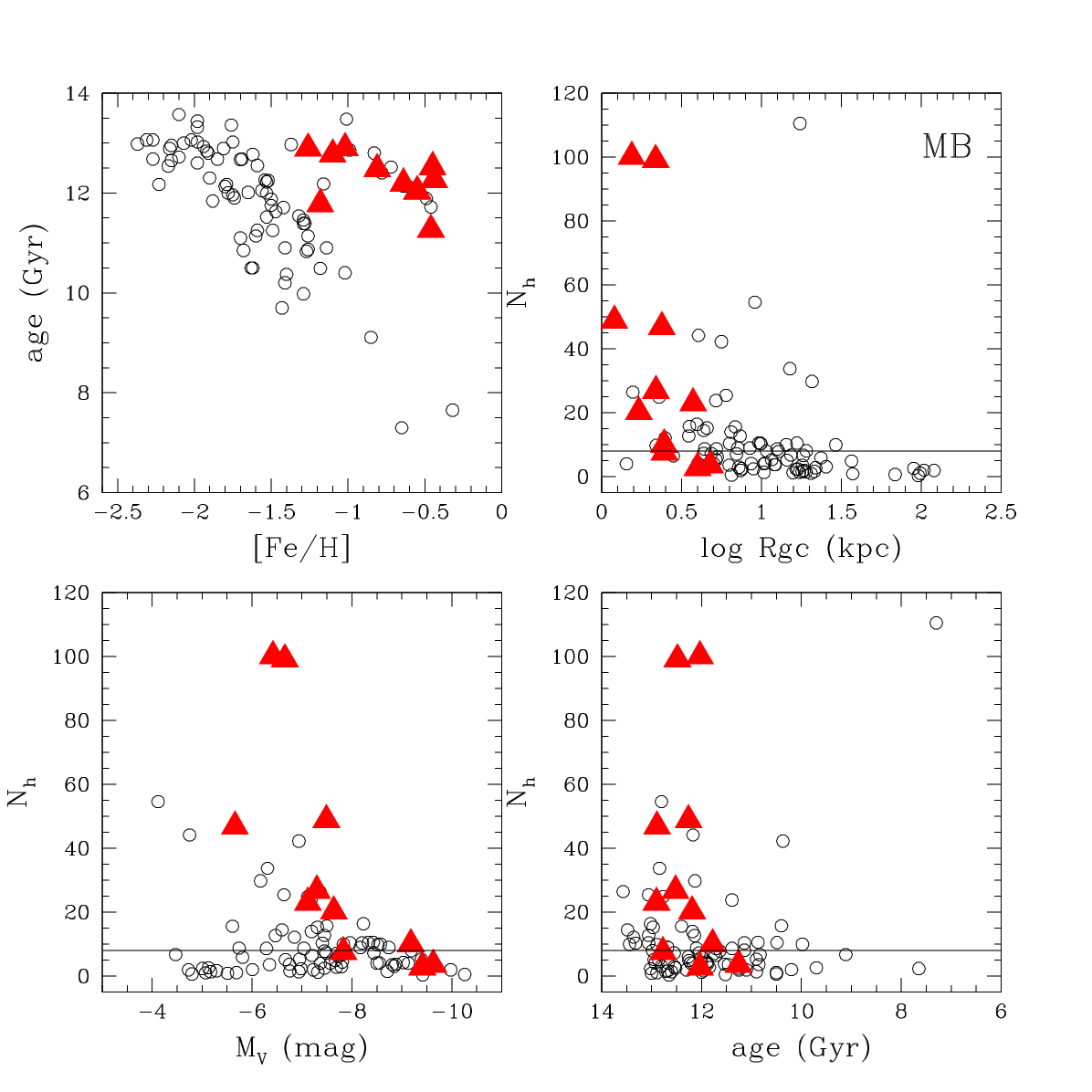}
\includegraphics[scale=0.40]{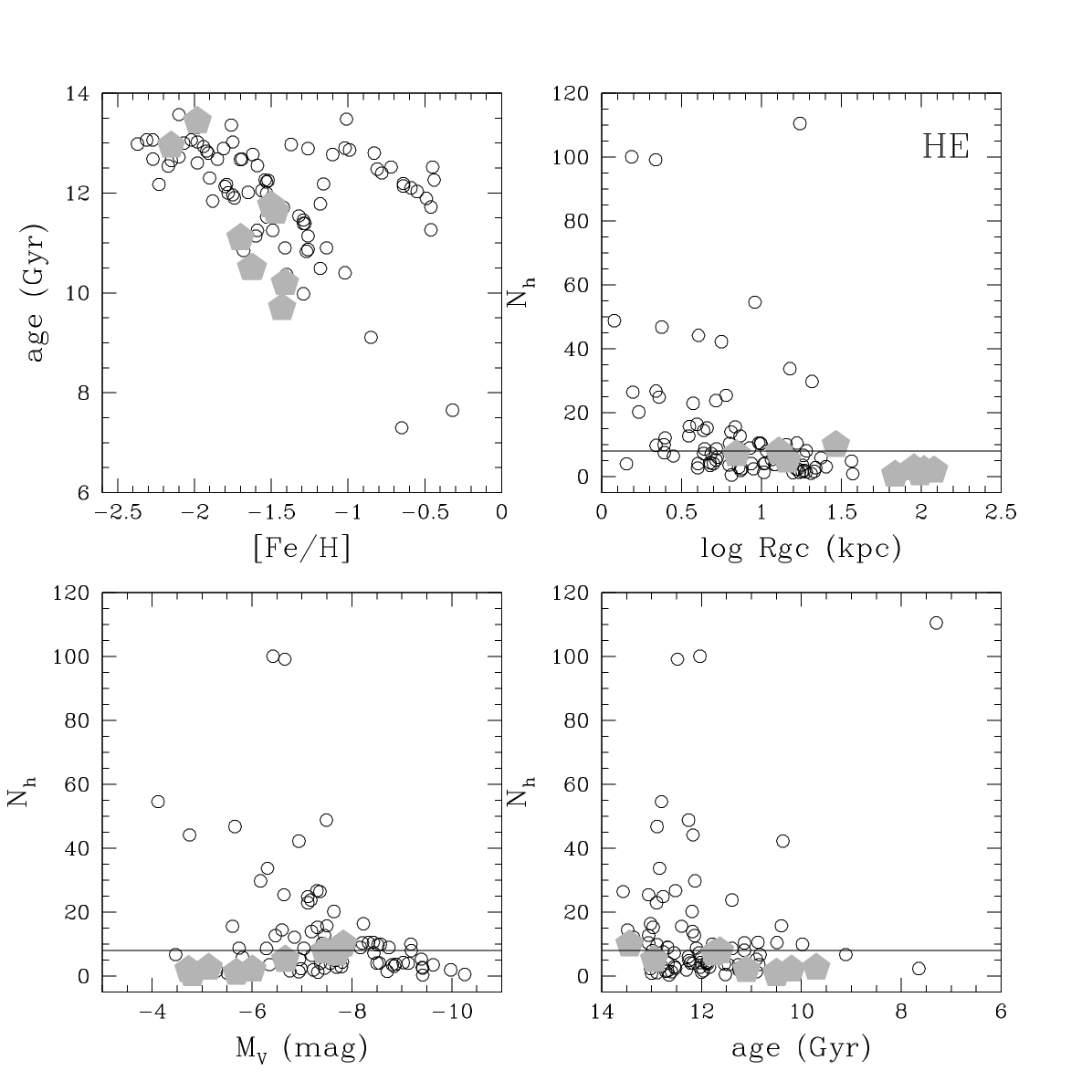}\includegraphics[scale=0.40]{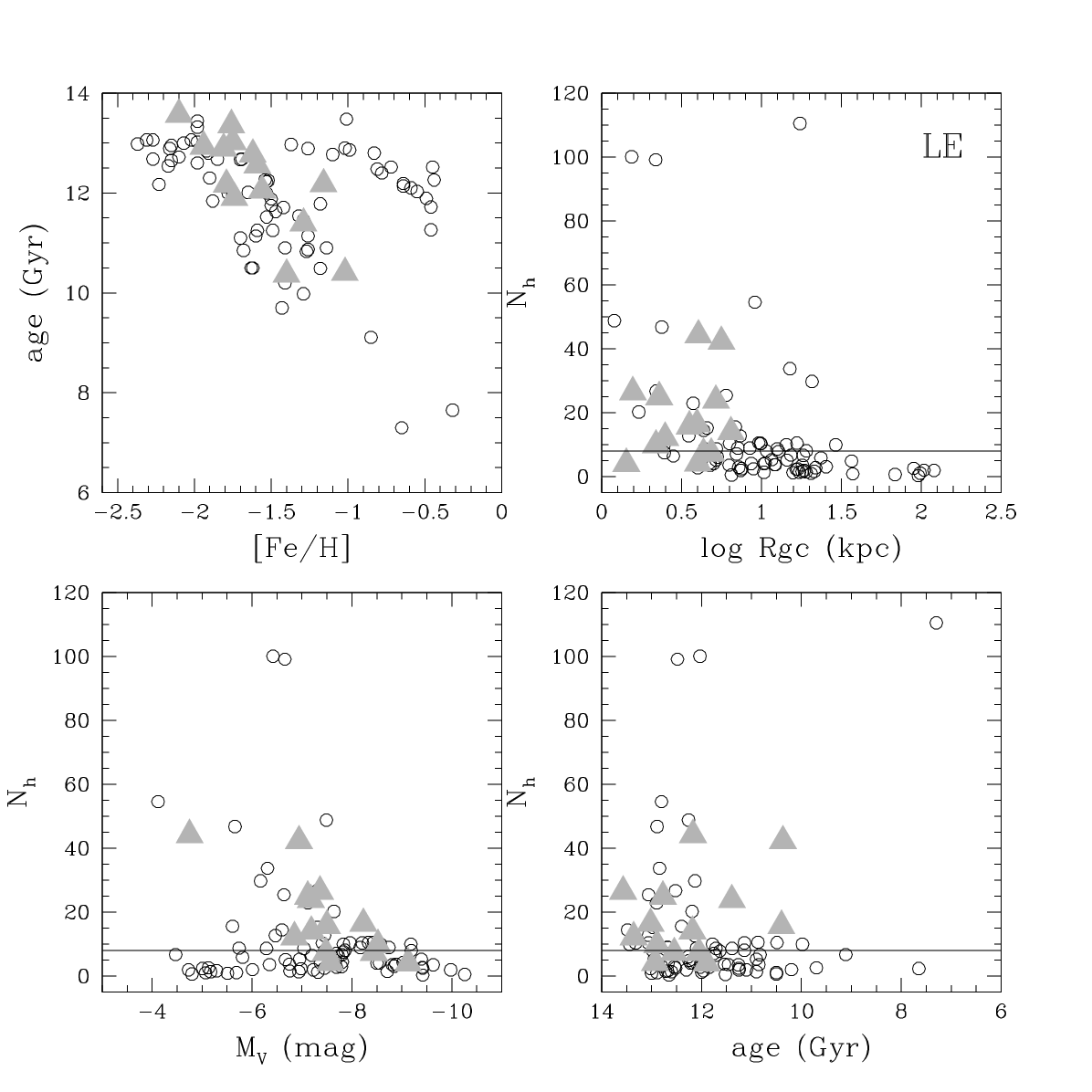}
\caption{Age-metallicity relation (upper left panel) and dynamical ages as a
function of the Galactocentric distance (upper right panel), total absolute
magnitude (lower left panel), and age (lower right panel) for GCs of the MD, MB,
HE, and LE groups. Empty small circles indicate the whole sample of GCs, 
larger symbols as in Fig.~\ref{f:nhfe} show GCs with different origin. The
horizontal line is traced at $N_h=8$.}
\label{f:appfig1}
\end{figure*}

\begin{figure*}[h]
\centering
\includegraphics[scale=0.40]{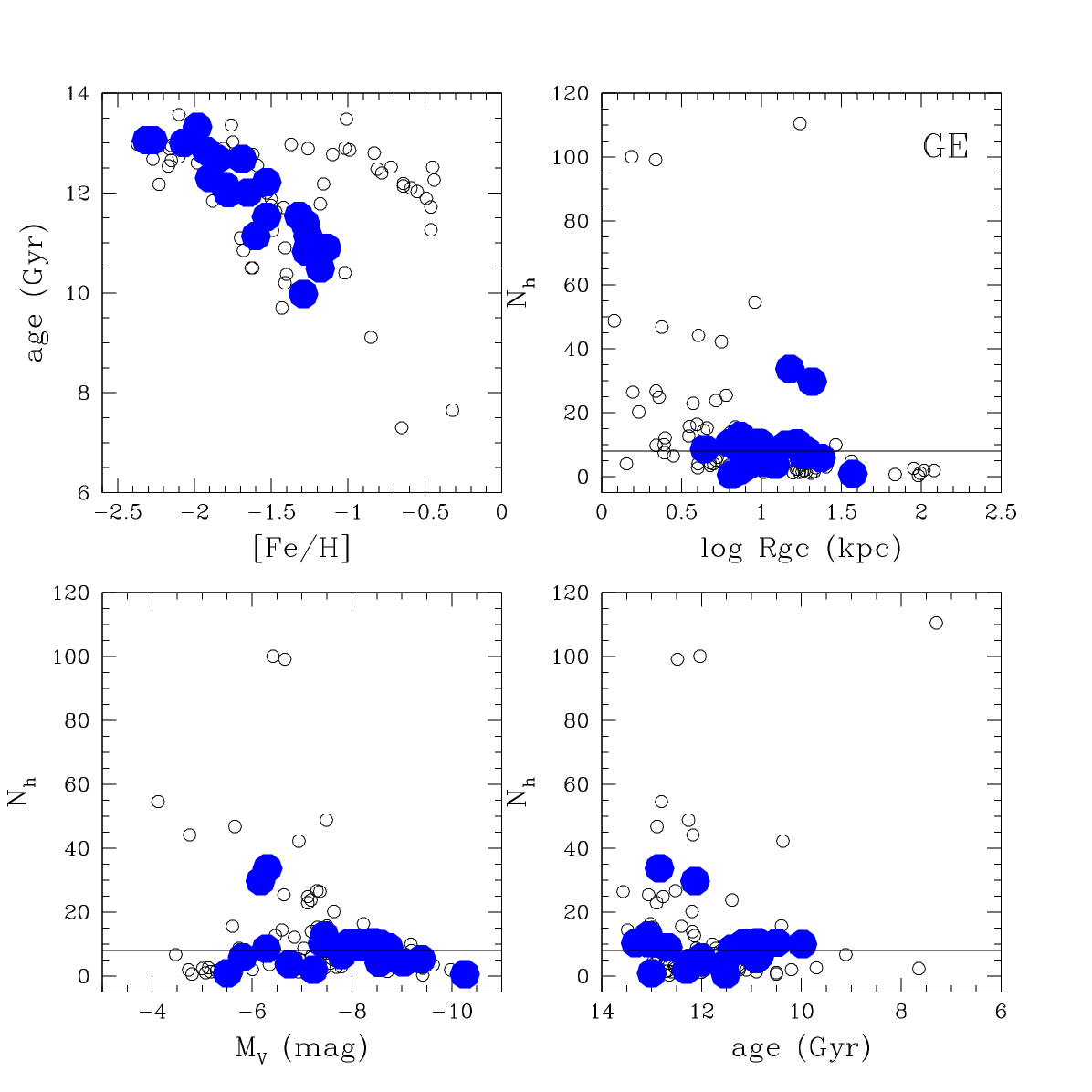}\includegraphics[scale=0.40]{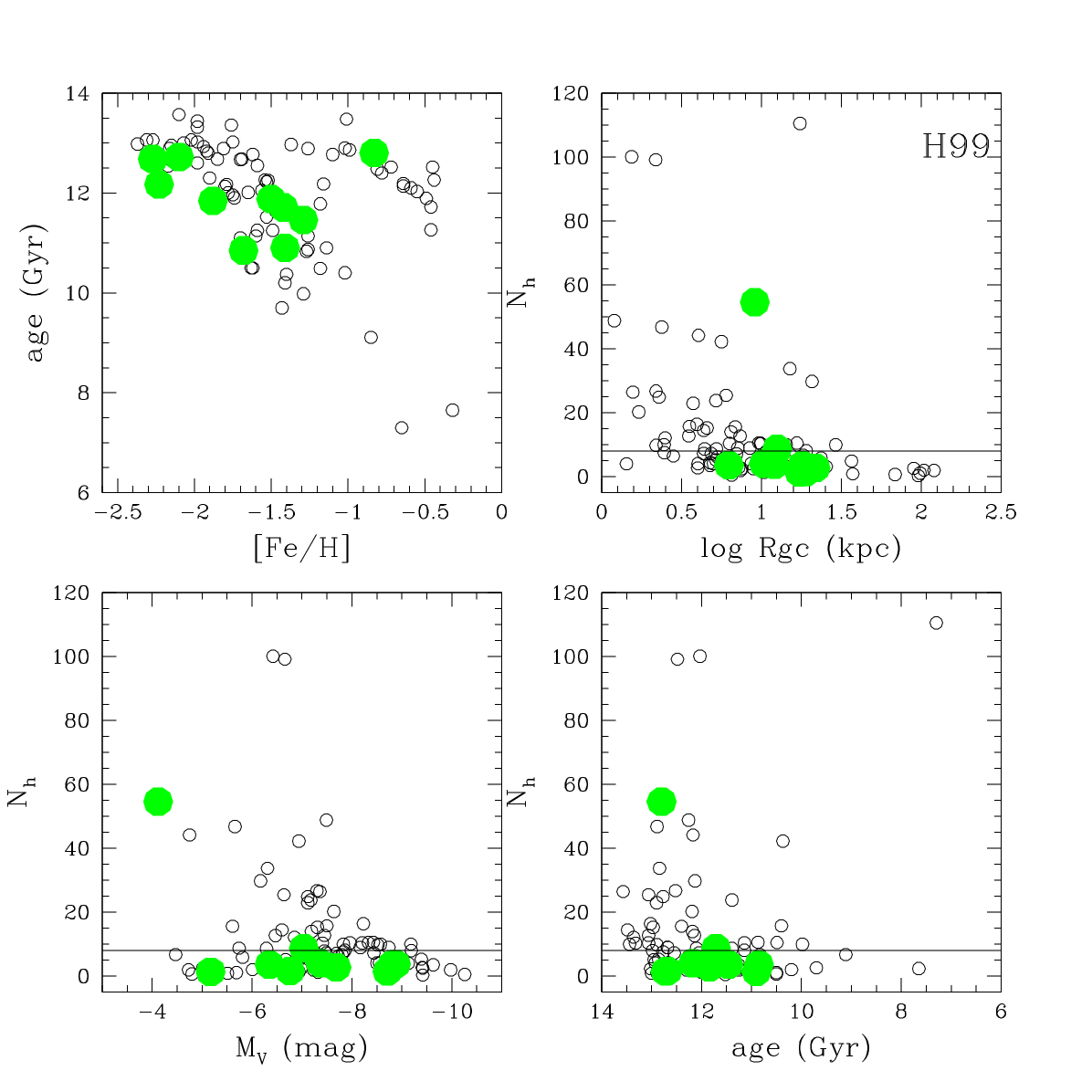}
\includegraphics[scale=0.40]{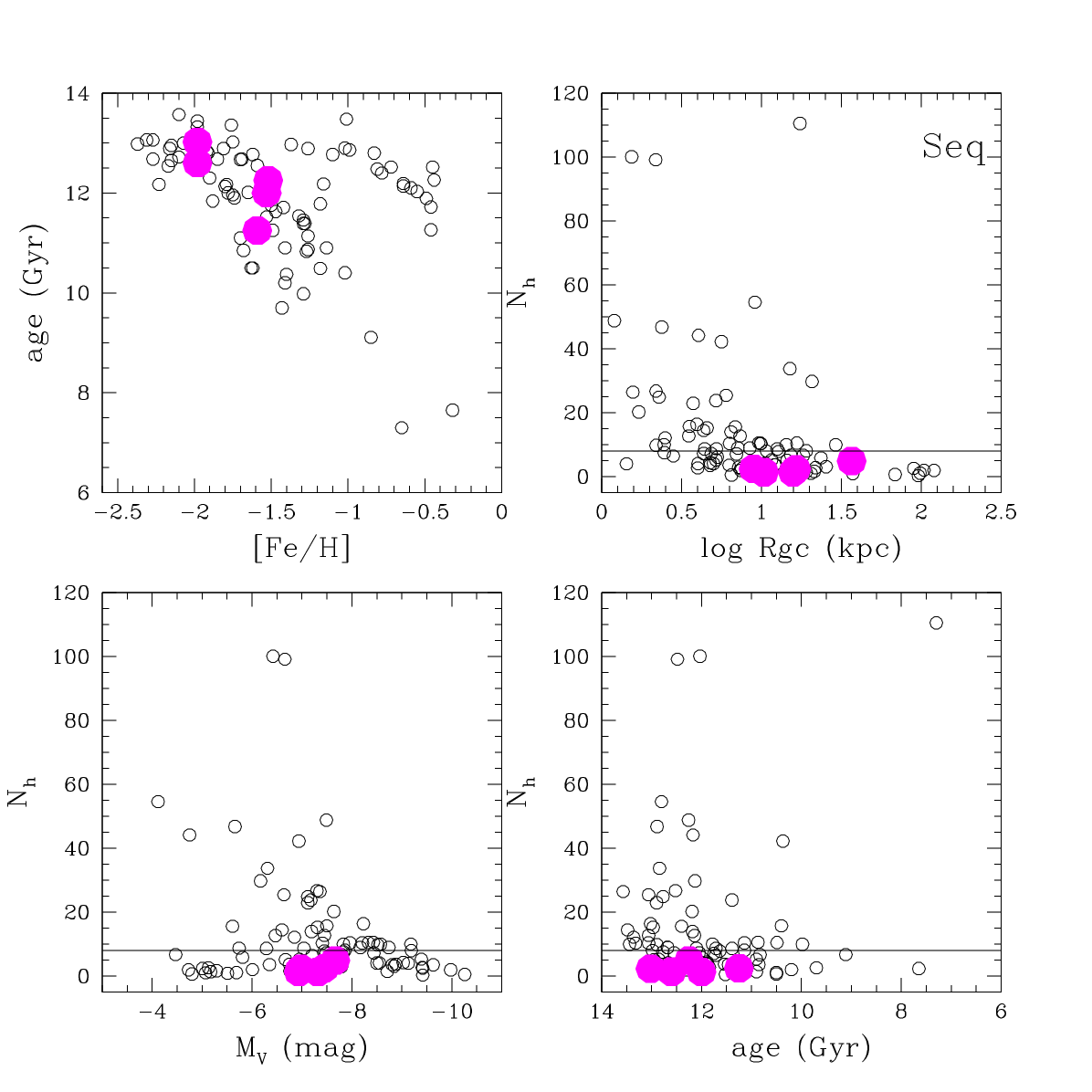}\includegraphics[scale=0.40]{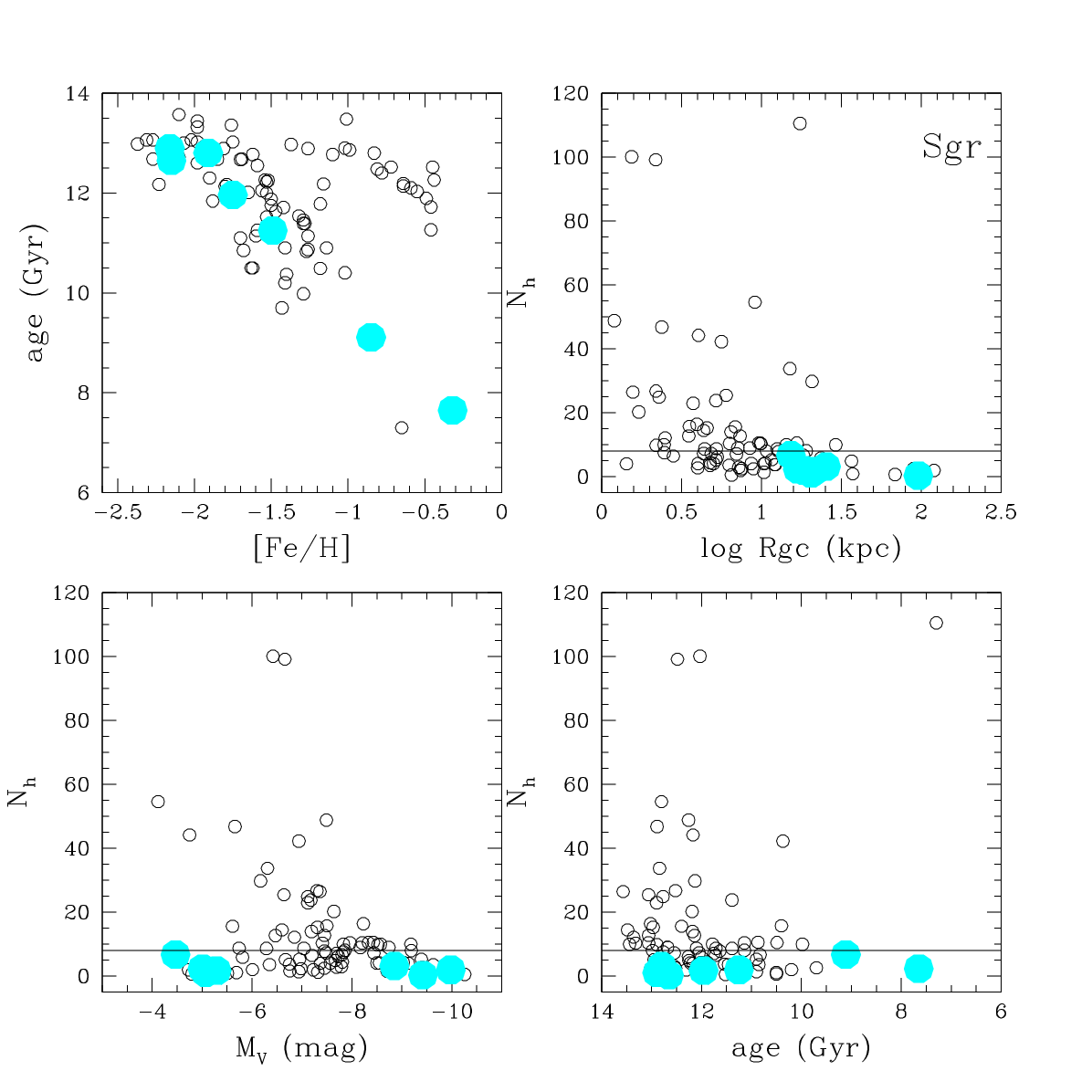}
\caption{As in Fig.~\ref{f:appfig1} for GCs of the GE, H99, SEQ, and SGR
groups. Symbols are as in Fig.~\ref{f:nhfe}.}
\label{f:appfig2}
\end{figure*}

\end{appendix}
\end{document}